%% file: mpc_est_journal_twocol_final.tex
\documentclass[dvipdfm,journal]{IEEEtran}
\IEEEoverridecommandlockouts

\usepackage{cite}

\usepackage{amsmath,amssymb,amsfonts}
\usepackage{algorithmic}
\usepackage{textcomp}
\def\BibTeX{{\rm B\kern-.05em{\sc i\kern-.025em b}\kern-.08em
    T\kern-.1667em\lower.7ex\hbox{E}\kern-.125emX}}
\usepackage{tabularx}
\usepackage[pdftex]{graphicx}
\usepackage{epstopdf}

\input{akbar_defs_new.tex}

\newcommand{\T}{{{\mbox{\tiny{T}}}}}

\newcommand{\A}{{\tiny{\mbox{A}}}}
\newcommand{\E}{{\tiny{\mbox{E}}}}
\newcommand{\W}{{\tiny{\mbox{W}}}}

\newcommand{\tdagger}{{\tiny{\dagger}}}
\newcommand{\tDelta}{{\tiny{\Delta}}}

\newcommand{\x}{{\rm{x}}}
\newcommand{\y}{{\rm{y}}}

\newcommand{\fov}{{\mbox{FoV}}}
\newcommand{\rot}{{\tiny{\mbox{rot}}}}
\newcommand{\LS}{{\tiny{\mbox{LS}}}}

\newcommand{\N}{{\tiny{\mbox{N}}}}
\newcommand{\Nx}{{\N_{\tiny{\mbox{x}}}}}
\newcommand{\Ny}{{\N_{\tiny{\mbox{y}}}}}

\newcommand{\alphah}{{\hat{\alpha}}}
\newcommand{\phA}{{\phi^{\A}}}
\newcommand{\phE}{{\phi^{\E}}}

\newcommand{\muh}{{\hat{\mu}}}

\begin{document}

\title{A Framework for Developing and Evaluating Algorithms for Estimating Multipath Propagation Parameters from Channel Sounder Measurements}

\author{\IEEEauthorblockN{\textsuperscript{1}Akbar Sayeed, \textsuperscript{2}Damla Guven, \textsuperscript{3}Michael Doebereiner, \textsuperscript{3}Sebastian Semper,  \textsuperscript{4}Camillo Gentile, \textsuperscript{4,5}Anuraag Bodi, \textsuperscript{6}Zihang Cheng}
\and \\[2mm]
\begin{tabular}{ll}
\begin{minipage}{2.8in}
\IEEEauthorblockA{\textsuperscript{1}\textit{\small Independent Researcher and Consultant}\\
\small{Madison, WI, USA}\\
\small{akbarmsayeed@gmail.com}}
\end{minipage}
&  
\begin{minipage}{3.4in}
\IEEEauthorblockA{\textsuperscript{2}\textit{\small Associate, Radio Access \& Propagation Metrology Group}\\
\textit{\small National Institute of Standards and Technology}\\
\small{Gaithersburg, MD, USA}\\
\small{damla.guven@nist.gov}}
\end{minipage}
\end{tabular}
\and \\[2mm]
\begin{tabular}{ll}
\begin{minipage}{2.8in}
\IEEEauthorblockA{\textsuperscript{3}\textit{\small TU Ilmenau}\\
\small{Ilmenau, Germany} \\
\small{michael.doebereiner@tu-ilmenau.de} \\
\small{sebastian.semper@tu-ilmenau.de}}
\end{minipage} & 
\begin{minipage}{3.4in}
\IEEEauthorblockA{\textsuperscript{4}\textit{\small Radio Access \& Propagation Metrology Group}\\
\textit{\small National Institute of Standards and Technology}\\
\small{Gaithersburg, MD, USA}\\
\small{\{camillo.gentile, anuraag.bodi\}@nist.gov}}
\end{minipage}
\end{tabular}
\and \\[2mm]
\begin{tabular}{ll}
\begin{minipage}{2.8in}
\IEEEauthorblockA{\textsuperscript{5}\textit{\small Contractor, Prometheus Computing}\\
\small{Gaithersburg, MD, USA}\\
\small{anuraag.bodi@nist.gov}}
\end{minipage}
&
\begin{minipage}{3.4in}
\IEEEauthorblockA{\textsuperscript{6}\textit{\small Electrical \& Computer Engineering}\\
\textit{\small U. Southern California}\\
\small{Los Angeles, CA, USA}\\
\small{zihangch@usc.edu}}
\end{minipage}
\end{tabular} 
}

\maketitle


\newpage

\begin{abstract}
A framework is proposed for developing and evaluating algorithms for extracting multipath propagation components (MPCs) from measurements collected by channel sounders at millimeter-wave frequencies. Sounders equipped with an omni-directional transmitter and a receiver with a uniform planar array (UPA) are considered. An accurate mathematical model is developed for the spatial frequency response of the sounder that incorporates the non-ideal cross-polar beampatterns for the UPA elements. Due to the limited Field-of-View (FoV) of each element, the model is extended to accommodate multi-FoV measurements in distinct azimuth directions. A beamspace representation of the spatial frequency response is leveraged to develop three progressively complex algorithms aimed at solving the single-snapshot maximum likelihood estimation problem: greedy matching pursuit (CLEAN), space-alternative generalized expectation-maximization (SAGE), and RiMAX. The first two are based on purely specular MPCs whereas RiMAX also accommodates diffuse MPCs. Two approaches for performance evaluation are proposed, one with knowledge of ground truth parameters, and one based on reconstruction mean-squared error. The three algorithms are compared through a demanding channel model with hundreds of MPCs and through real measurements. The results demonstrate that CLEAN gives quite reasonable estimates which are improved by SAGE and RiMAX. Lessons learned and directions for future research are discussed.
\end{abstract}

\vspace{-3mm}
\section{Introduction}
\label{sec:intro}
\begin{figure*}[htb]
\centering
\includegraphics[width=7.0in]{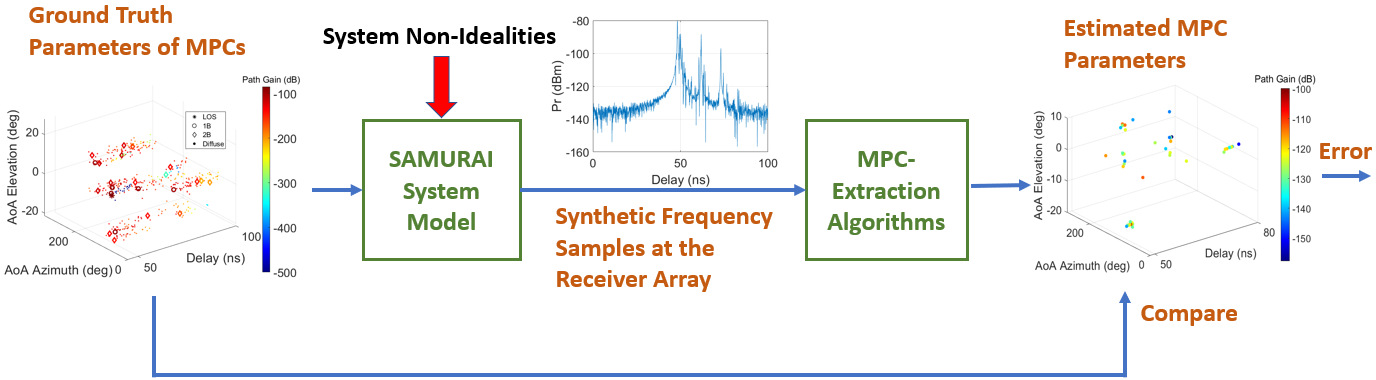}
\vspace{-3mm}
\caption{\footnotesize{\sl Illustration of the proposed methodology for the development and evaluation of MPC parameter extraction algorithms.  A mathematical model for sounder, incorporating  non-idealities, is used to synthetically generate the sounder output - the space-frequency response of the propagation environment - for a given set of known MPC parameters, which serve as  the ``ground truth''.  The synthetic space-frequency channel response serves as the input to the MPC estimation algorithms, that also leverage the sounder model.}}
\label{fig:method}
\vspace{-3mm}
\end{figure*}
Accurate modeling of the multipath propagation environment is critical for the design and deployment of wireless networks, especially at millimeter-wave (mmWave) and Terahertz (THz) frequencies that are part of 5G, 6G and emerging standards because of the inherently higher spatial and temporal resolutions at these frequencies. Accurate channel modeling in turn relies on appropriate and precise measurements of the propagation environment collected by radio-frequency channel sounders. Wideband directional sounders at mmWave and THz frequencies can take on various forms depending on the beamforming mechanism, such as phased arrays, lens arrays or mechanically pointed horn antennas  \cite{ch_ams:brady_taps:12, ch_ams:sayeed_gcom:16}, which in turn dictates different approaches for measuring, calibrating and modeling sounder characteristics. The performance of the multipath propagation component (MPC) extraction depends on both the sounder hardware characteristics as well as the estimation algorithms used for processing the measurements. 

Building on initial work based on an ``idealized'' model for the National Institute of Standards and Technology (NIST) SAMURAI\footnote{Synthetic Aperture Measurement UnceRtainty for Angle of Incidence.} sounder \cite{sayeed_mpc:gcom20}, this paper details a new and comprehensive framework for the development and evaluation of MPC estimation algorithms, leveraging a realistic model for the sounder measurements that fully incorporates the measured cross-polar and frequency-dependent beampatterns for array elements in addition to measurement noise. The non-ideal measured beampatterns and the inherent forward-backward ambiguity also necessitates a new formulation for multiple Field-of-Views (FoVs) to accurately characterize the entire 360 degree FoV in azimuth. These new contributions reported in this paper are unprecedented and reflect the culmination of a multi-year effort\footnote{It reflects the completed work of two sub-groups from the Alliance, one focused on the measurement aspects and one focused on the algorithmic aspects; Sayeed actively participated in both the sub-groups.} as part of ongoing work by the NIST NextG Channel Model Alliance, including a recent work on benchmarking the performance of sounders \cite{nist_twc:20}. The new contributions are reflected both in the new system (sounder and channel) model as well as in the development of the algorithms for processing the measurements.

\vspace{-3mm}
\subsection{Overview of the Methodology}
\label{sec:overview}

The proposed framework for developing and evaluating MPC extraction algorithms is illustrated in Fig.~\ref{fig:method}.  The measurements collected by the sounder are processed by the algorithms to estimate the MPCs of the propagation channel.  In the proposed evaluation of estimation algorithms, the actual channel measurements are replaced with a realistic mathematical model for the sounder as illustrated in Fig.~\ref{fig:method}.  A known set of MPC parameters (``ground truth'' (GT)) is provided as an input to the sounder model which then generates the corresponding spatial frequency response matrix as the output.  The MPC parameter estimation algorithms, that incorporate the realistic sounder model, are then applied to the spatial frequency response matrix generated by the sounder model. A statistical analysis of the estimated MPC parameters, relative to the GT values, is provided to evaluate the performances of different algorithms. Another criterion, based on the normalized mean-squared reconstruction error, that does not require GT data, is also considered. 

While the framework developed is general, the special case considered in this paper is the single input multiple output (SIMO) setting in which the TX of the sounder is assumed to have a single isotropic antenna, and the RX is equipped with a uniform planar array (UPA), as in the NIST SAMURAI system, with a WR-28 waveguide serving as the UPA element. A UPA at the RX is sufficient to resolve the MPCs in the 3D azimuth (AZ), elevation (EL), and delay space; including a UPA at the TX would have significantly increased computational complexity. A realistic model for the SAMURAI system is developed to enable an extensive evaluation of the estimation algorithms based on synthetic measurements generated from the NIST GT MPC data. The results on the performance of the MPC estimation algorithms for real measurements provide additional cross-validation of the algorithms developed and evaluated using the synthetic MPC data. 

The development in this paper falls under the ``narrowband scenario'' where the bandwidth (1 or 2 GHz) is small compared to the operating frequency (28 GHz) so that the array response vectors are assumed to be invariant over the bandwidth \cite{brady:icc15}. However, the measured  beampatterns used for each WR-28 UPA element are frequency dependent. Three algorithms aimed at solving the single-snapshot maximum-likelihood (ML) estimation problem are considered: CLEAN, SAGE, and RiMAX. CLEAN and SAGE assume a specular MPC model, and RiMAX incorporates a diffuse MPC model as well.

\begin{figure}[hbt]
\vspace{-2mm}
\begin{tabular}{c}
\hspace{-3mm}\includegraphics[width=3.2in]{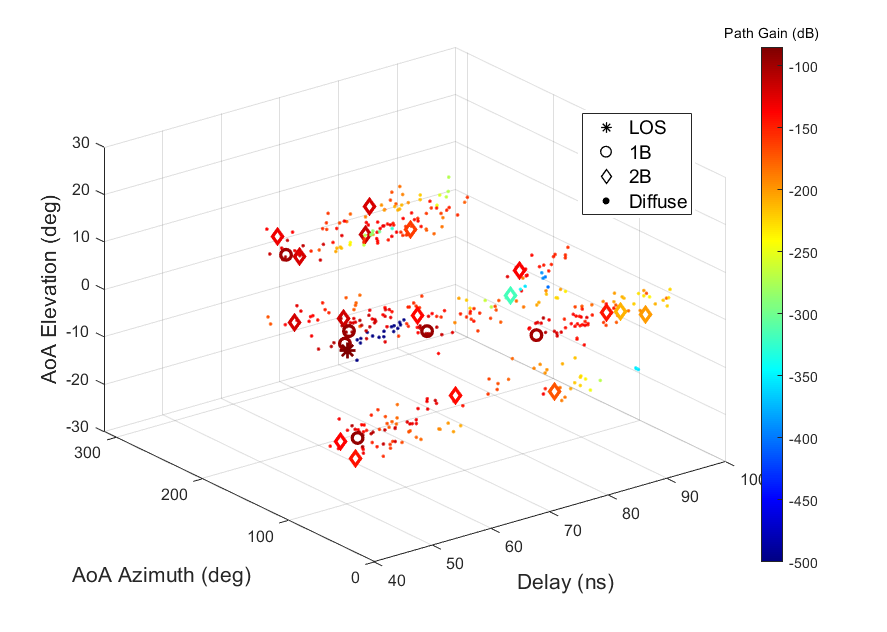} \\
\footnotesize{(a)} \\
\hspace{-3mm}\includegraphics[width=3.2in]{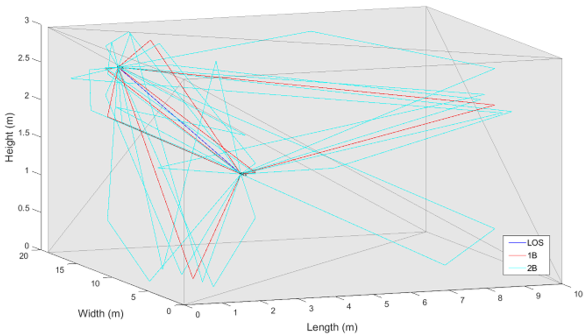} \\
 \footnotesize{(b)} 
\end{tabular}
\vspace{-1mm}
\caption{\footnotesize{\sl Illustration of the GT MPC data for one TX-RX location. (a) A 3D plot of LoS, single bounce (1B), double bounce (2B), and diffuse MPCs. The diffuse MPCs are generated statistically, clustered around each 1B and 2B MPC. (b) The rays corresponding to specular MPCs.}}
\label{fig:gt_data}
\vspace{-3mm}
\end{figure}
The GT data used in this paper is provided by NIST \cite{nist_mpc:19} based on propagation measurements, over a wide frequency range (13.5 GHz to 40 GHz), in a conference room of dimension 10 m $\times$ 19 m $\times$ 3 m, for 10 distinct TX-RX locations, as shown in Fig.~\ref{fig:gt_data}. Fig.~\ref{fig:gt_data}(a) shows an illustrative example for one TX-RX location, featuring over 450 MPCs including the direct path, first- and second-order specular reflections, and diffuse components that are clustered around each specular MPC \cite{nist_mpc:19}. Fig.~\ref{fig:gt_data}(b) shows the ray-traced model for the specular paths. For each TX-RX location, three sounder measurements are synthetically generated, corresponding to the RX array pointing in three directions, centered on disjoint  120-degree sectors in azimuth.

\vspace{-3mm}
\subsection{Related Work}
\label{sec:prior_work}
Extensive research has been conducted to develop and compare the performance of MPC estimation algorithms, which can be broadly categorized into three main groups: spectral subspace-based techniques, parametric subspace-based methods, and  parametric ML-based  methods  \cite{riu}, \cite{krim}. Multiple Signal Classification (MUSIC) and Unitary estimation of signal parameter via rotational invariance techniques (ESPRIT) belong to the first two groups, respectively, while CLEAN, SAGE and RiMAX are part of the final category \cite{riu, krim}. Subspace-based methods assume knowledge of second-order channel statistics, which necessarily requires multiple statistically uncorrelated (in the MPC amplitudes) measurements, or snapshots, so that the resulting covariance matrix of the measurement data exhibits full rank in the signal subspace \cite{riu,krim}. On the other hand, ML-based parametric approaches, as those considered in this paper, can be applied to single-snapshot measurements, which are more feasible in practice. For this reason, single-snapshot approaches are also sometimes referred to as ``deterministic'' and covariance-based multiple-snapshot approaches as ``stochastic.'' One technique for artificially creating multiple snapshots from a single snapshot for subspace methods is ``smoothing'': the single snapshot is partitioned, possibly with some overlap, in the spatial or frequency domain, to generate multiple snapshots from which an estimate for the covariance matrix in the corresponding domain can be obtained by averaging, albeit at a loss of angular or delay resolution.

Another difference between various MPC parameter estimation approaches is whether the parameters are estimated jointly (as in MUSIC and ESPRIT, typically) or sequentially (iteratively) as in CLEAN, SAGE and RiMAX. These differences in attributes also influence which algorithms are evaluated in the same setting. Performance criteria considered in the comparison of algorithms include the accuracy in estimating the MPC parameters (e.g., azimuth, elevation, time delay), applicability to different array architectures, convergence speed and computational complexity, as discussed in the review papers  \cite{riu,krim}. Studies in \cite{gaillot, tanghe} report performance results on RiMAX, SAGE and ESPRIT and investigate the impact of including diffuse multipath components (DMCs) in the model which is only accomplished by RiMAX. The comparison of SAGE and ESPRIT is based on a model with only specular multipath components (SMCs). The findings highlight the importance of DMCs, as RiMAX is shown to yield the best performance \cite{gaillot,tanghe}. It is also found that SAGE returns more accurate angular estimates than ESPRIT \cite{tanghe}. In other related work, MPC estimation algorithms are tested under simulation scenarios with varying array sizes and signal-to-noise ratio (SNR) \cite{riu,feng,yin,damico}. Although SAGE is reported to have higher computational complexity due to its iterative nature, compared to MUSIC and ESPRIT, it performs better under low-SNR scenarios and is more robust to channel modeling errors \cite{yin}, \cite{yao}. The CLEAN algorithm is shown to be an effective MPC estimation algorithm in the single-snapshot (deterministic) setting, however comparison with other techniques is rare \cite{naitong}.

Generally, most existing studies analyze the estimation techniques via Monte-Carlo simulations based on simplified synthetic scenarios limited to first-order specular components with less than ten MPCs \cite{riu, feng, yin, damico, yao, tschudin}. Studies that report results on real measurements are generally limited with little or no knowledge of the ground truth \cite{riu, krim, feng, yin, naitong, tschudin}. Relatively few papers evaluate algorithms in both synthetic and real settings \cite{riu, feng, yin, tschudin}. However, it is crucial to test the algorithms in presence of practically relevant multipath mechanisms, e.g., first- and higher-order specular paths, and diffuse components, in addition to the direct path, as a more realistic representation of a propagation environment.  It is also critical that the performance of algorithms is investigated with a realistic channel sounder model which incorporates system non-idealities, such as antenna element beampatterns. However, none of the studies thus far have accomplished that.

Table~\ref{fig:comp_table} summarizes the key attributes of the related work reviewed above in the context of the contributions of this paper. In particular, none of the related works address three key innovations in the mathematical framework developed in this paper: modeling of non-ideal antenna element beampatterns, fully polarimetric modeling of measurements, and a multi-FoV formulation, both for simulating the measurement data and for algorithm design.

\begin{table*}[hbt]
\vspace{-1mm}
\centerline{\includegraphics[width=7.0in]{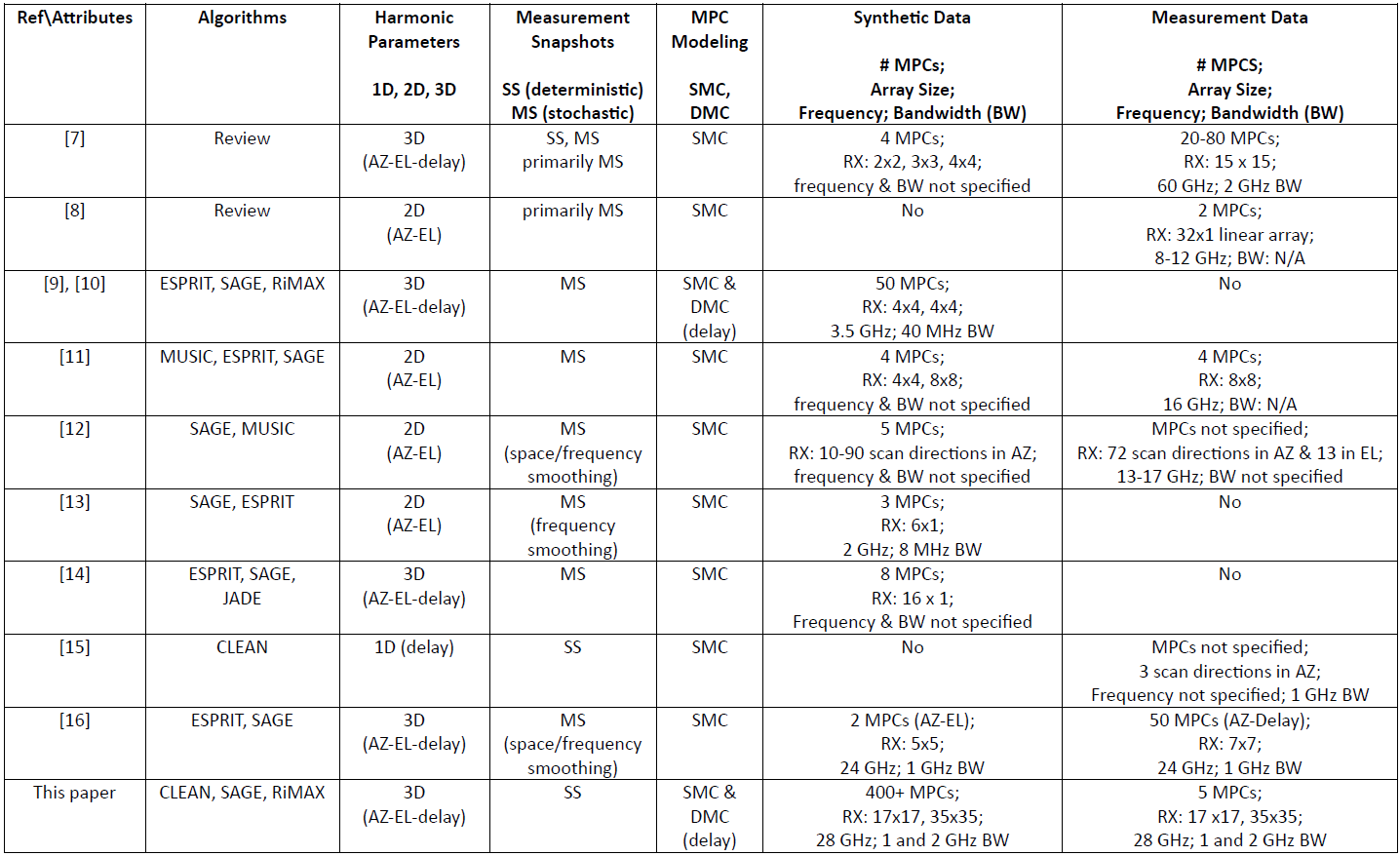}} 
\caption{\footnotesize{\sl A table summarizing the key attributes of related work in the context of this paper. None of the related works address three key innovations in the mathematical framework developed in this paper: modeling of non-ideal antenna element beampatterns, fully polarimetric modeling of measurements, and a multi-FoV formulation, both for simulating measurement data and for algorithm design. Furthermore, unlike related work, the measurement results reported in this paper are for the NIST SAMURAI system in a controlled environment with accurate measurements of the harmonic parameters of the strong MPCs.}}
\label{fig:comp_table}
\vspace{-4mm}
\end{table*}

\vspace{-3mm}
\subsection{Summary of Novel Contributions}
\label{sec:contributions}
The focus of this paper is on deterministic, single-snapshot parametric MPC estimation algorithms: CLEAN, SAGE and RiMAX. Subspace-based methods, such as MUSIC and ESPRIT are not directly applicable in this practically relevant setting. The proposed framework develops the new mathematical model and description for the three algorithms in an integrated and complete fashion and as a result provides an explicit blueprint for implementing the algorithms. The mathematical development is supported by the numerical results on the implementation in the most extensive simulation setting based on previously estimated MPCs at NIST as well as real measurements taken by the actual SAMURAI system (which is accurately captured by the mathematical model developed in this paper) in a representative and carefully designed environment.

Additional significant and novel contributions of the framework developed in this paper include:
\begin{itemize}
\item A realistic model for the sounder measurements, developed in Sec.~\ref{sec:nonideal}, that incorporates the non-ideal and frequency-dependent beampatterns of UPA elements in addition to measurement noise. The model is extended to a fully polarimetric setting in Sec.~\ref{sec:polarization} and measured cross-polar beampatterns for the UPA elements (WR-28 waveguide) are utilized in the numerical results in Sec.~\ref{sec:results}.
\item An integrated new development of the three algorithms in Sec.~\ref{sec:extraction}, encompassing both specular MPCs (CLEAN and SAGE) and diffuse MPCs (RiMAX), that incorporates the new realistic system (channel and sounder) model in algorithm design.  
\item A multi-FoV formulation developed in Sec.~\ref{sec:fov}, necessitated by the non-ideal element beampatterns, in which measurements from three UPA orientations are jointly processed for MPC extraction. The multi-FoV formulation encompasses both the system (channel and sounder) model as well as algorithm design.
\item Identification of two appropriate metrics in Sec.~\ref{sec:metrics} for evaluating the MPC extraction algorithms: one that relies on the knowledge of ground truth MPC parameters for insight, and one based on the normalized mean-squared reconstruction error that is more appropriate in practice. An analysis of the computational complexity of the three algorithms is also provided.
\item Evaluation of the algorithms in Sec.~\ref{sec:results} for two different UPA sizes (17$\times$17, 35$\times$35) and two different bandwidths (1, 2 GHz) at a center frequency of 28 GHz, to test the robustness of the algorithms with array size and bandwidth. In particular, the larger array size (35 $\times$ 35) and bandwidth (2 GHz) are approaching the boundary of the ``narrowband'' scenario \cite{brady:icc15} as noted in the results.  
\item Evaluation of the algorithms using both synthetic and real measured data. The synthetic evaluation in Sec.~\ref{sec:syn_meas} is based on one of the most extensive and accurate MPC data sets, consisting of over 400 MPCs with both specular MPCs (direct path and first-order and second-order reflections) and diffuse MPCs for 10 different TX-RX locations. The real measurements used in the evaluation in Sec.~\ref{sec:measurements} were taken by the NIST SAMURAI sounder system in a carefully designed scenario aimed at testing the resolution capabilities of the sounder with precise and accurate measurements of the ground truth MPCs.  
\end{itemize}

\noindent
{\bf Notation:} The superscript $^\T$ refers to transpose, $^*$ to complex conjugation, and $^\tdagger = ^{\T *}$ to complex conjugate transpose.  

\vspace{-3mm}
\section{Physical Channel Model and Its Beamspace Representation}
\label{sec:chan_mod}
This section describes an idealized physical model for the multipath propagation channel for a UPA and its sampled beamspace representation, induced by key sounder parameters. The model is extended to a realistic model incorporating hardware non-idealities in Sec.~\ref{sec:nonideal} and to multiple FoVs in Sec.~\ref{sec:fov}.

\vspace{-3mm}
\subsection{Physical SIMO Model for A Uniform Planar Array}
\label{sec:phy_ams}
In the static scenario, the physical model can be expressed as a spatial frequency response  \cite{ch_ams:bonek:01,ch_ams:cost2100:12,ch_ams:sayeed:02,ch_ams:sayeed_book:08}
\begin{align}
\bH(f) & = \sum_{n=1}^{N_p} \alpha_n \ba_{\Nx}(\theta^\x_{n})  \ba_{\Ny}^{\tdagger}(\theta^\y_{n}) e^{-j2\pi \tau_n f} 
\label{Hf} 
\end{align}
which represents a SIMO channel connecting an omni-directional single-antenna TX to a RX UPA with omni-directional antennas, $N_\x$ in the $x$ (horizontal) direction and $N_\y$ in the $y$ (vertical) direction. The channel is represented by the  $N_\x \times N_\y$  spatial frequency response matrix $\bH(f)$  which captures the signal propagation over $N_p$ paths, with $\alpha_n$,  $\theta^\x_n$,  $\theta^\y_n$, and $\tau_n$  denoting the complex amplitude, spatial frequency in x direction, spatial frequency in the y-direction, and delay of the $n$-th path.  The spatial frequencies  in (\ref{Hf}) are induced by the physical angles in azimuth (AZ) and elevation (EL), $\phi^\A_{n} \in [-\pi,\pi)$ and  $\phi^\E_{n} \in [-\frac{\pi}{2},\frac{\pi}{2}]$, defined with respect to the broadside direction, via the relationship: 
\begin{equation}
\theta^\x  =\frac{d_\x}{\lambda} \sin(\phA) \cos(\phE) \ ; \ \theta^\y  = \frac{d_\y}{\lambda} \sin(\phE) \ ,
\label{th}
\end{equation}
where $d_\x$ and $d_\y$ are the antenna spacings in the x and y directions, and $\lambda$ is the operating wavelength. We assume that $d_\x= d_\y = d$. The vector  
$\ba_{\Nx}(\theta^\x)$ is an $N_\x \times 1$ response vector for the x-direction and $\ba_{\Ny}(\theta^\y)$ is an $N_\y \times 1$ response vector for the y-direction. 
The response vectors take the form of discrete spatial sinusoids with frequencies $\theta^\x, \theta^\y$  \cite{ch_ams:bonek:01,ch_ams:sayeed:02,ch_ams:sayeed_book:08}:
\begin{align}
\hspace{-3mm} \ba_{\N}(\theta) \! & = \! \left [ 1, e^{-j2\pi \theta}, \cdots, e^{-j2\pi \theta(N - 1)} \right ]^{\T}  , \ \theta \in \left [-\frac{d}{\lambda}, \frac{d}{\lambda} \right ] .
\label{st_vec}  
\end{align}
The relationship (\ref{th}) induces a one-to-one correspondence between $\theta^\y \in \left [-\frac{d}{\lambda}, \frac{d}{\lambda} \right ]$ and $\phi^\E \in \left [-\frac{\pi}{2}, \frac{\pi}{2} \right ]$. Similarly, there is a one-to-one correspondence between $\theta^\x \in \left [-\frac{d}{\lambda}, \frac{d}{\lambda} \right]$ and $\phi^\A \in \left [-\frac{\pi}{2},\frac{\pi}{2} \right]$ - the {\em principle range} of AZ angles. The AZ angles outside this range, get aliased back into it as 
\begin{equation}
\phi^\A \rightarrow -\pi - \phi^\A , \phi^\A \! \in \! \left [-\pi, -\frac{\pi}{2} \right )  ;  
\phi^\A \rightarrow \pi - \phi^\A , \phi^\A \! \in \! \left (\frac{\pi}{2}, \pi \right ) \ .
\nonumber
\end{equation}
The ambiguity between the forward and backward hemispheres requires multiple FoVs.

The model (\ref{Hf}) is widely used for simulating wireless channels. It assumes knowledge of the MPC parameters at perfect (infinite) angle-delay resolution. On the other hand, any sounder/system in practice has a finite resolution in angle-delay, which also impacts the statistical characteristics of the estimated MPC parameters  \cite{ch_ams:sayeed_book:08,ch_ams:brady_taps:12}. These challenges are accentuated at mmWave frequencies due to: i) the lack of sufficient measurements in different operational environments, and ii) limited capabilities of existing sounders, e.g., low spatial resolution and/or mechanical pointing. Fundamentally, many technical issues need to be addressed for estimating the angle-delay  MPC parameters from sounder measurements, especially for sounders with antenna arrays for directional measurements. The ``beamspace'' channel representation in angle-delay discussed next provides a useful tool for developing and comparing MPC extraction algorithms.

\vspace{-3mm}
\subsection{Beamspace Representation of the Physical Model}
\label{sec:sampled_ams}
A fundamental connection between the measurements made in practice and the physical model above is revealed by the beamspace representation of the physical model (\ref{Hf}) \cite{ch_ams:sayeed:02,ch_ams:sayeed_book:08}
\begin{align}
H_b(\theta^\x,\theta^\y,\tau)    = &   \frac{1}{\scriptstyle{W N_\x N_\y}}  \! \! \! \int_{\scriptstyle{-\frac{\W}{2}}}^{\scriptstyle{\frac{\W}{2}}}  \hspace{-2mm} \ba_{\Nx}^{\tdagger}(\theta^\x) \bH(f) \ba_{\Ny}(\theta^\y) 
  e^{j 2\pi \tau f} df \nonumber \\
= & \sum_{n=1}^{N_p} \alpha_n f_{\Nx}(\theta^\x - \theta^\x_n)  f_{\Ny}(\theta^\y -\theta^\y_{n}) 
\nonumber \\ &  
{\rm sinc}(W(\tau-\tau_n)) 
\label{Hb3d} 
\end{align}
where $W$ is the (two-sided) operational bandwidth and $H_{b}(\theta^\x, \theta^\y,\tau)$ represents the 3D channel impulse response in angle-delay space, ${\rm sinc}(x) = \frac{\sin(\pi x)}{\pi x}$ and $f_{\N}(\theta)$ denotes the Dirichlet sinc function $f_{\N}(\theta) = \frac{\sin(\pi N \theta)}{\pi N \theta}$. 
The beamspace representation maps the spatial frequency response in (\ref{Hf}) into the angle-delay (beamspace) domain through a 3D Fourier transform. A sampled version of $H_b(\theta^\x, \theta^\y, \tau)$ is used, directly or indirectly, in MPC estimation algorithms:
\begin{equation}
H_b[i,k,\ell] = H_b(i \tDelta \theta^\x, k \tDelta \theta^\y, \ell \tDelta \tau) \ .
\label{Hb3d_samp}
\end{equation} 
Critically spaced samples are defined by the temporal (delay) resolution and spatial resolution of the UPA in x and y directions at the RX:
\begin{equation}
\tDelta \tau = \frac{1}{W} \ ; \ \tDelta \theta^\x =  \frac{1}{N_\x} \ ; \ \tDelta \theta^\y = \frac{1}{N_\y} \ .
\label{samp_res}
\end{equation}
Critical spatial sampling induces an equivalent {\em beam-frequency} representation of $\bH(f)$ 
\[
\bH_b(f) = \bU_{\Nx}^{\tdagger} \bH(f) \bU_{\Ny} \Longleftrightarrow \bH(f)= N_\x N_\y \bU_{\Nx} \bH_b(f) \bU_{\Ny}^{\tdagger}   
\]
where $\bH_b(f)$ is the sampled beam-frequency response matrix, and the matrices $\bU_{\Nx}$ and $\bU_{\Ny}$ represent the spatial Discrete Fourier Transform (DFT) matrices, whose columns are steering/response vectors (\ref{st_vec}) for uniformly spaced spatial frequencies, that map the antenna domain into the angle (beam) domain:
$\bU_{\Nx} = \frac{1}{N_{\x}} \left [ \ba_{\Nx}(\tDelta \theta^\x), \ba_{\Ny}(2\tDelta \theta^\x), \cdots, \ba_{\Nx}(N_\x \tDelta \theta^\x) \right ]$ and similarly for $\bU_{\Ny}$. The sampled beam-frequency representation $\bH_b(f)$ is an equivalent representation  
of $\bH(f)$ over the bandwidth $W$ and contains all information about it.
The beamspace representation in angle-delay is particularly useful at mmWave frequencies due to the highly directional nature of propagation. It is a natural domain for representing channel measurements made with directional antennas; e.g., phased arrays, lens arrays or mechanically-pointed antennas \cite{ch_ams:sayeed_gcom:16}.

\vspace{-3mm}
\subsection{Physical Model with Antenna Element Beampatterns}
\label{sec:nonideal}
The ideal model in (\ref{Hf}) assumes ideal omni-directional beampatterns for each antenna element at the RX. In practice, each element has a particular antenna pattern that needs to be accounted for in the physical model. In this paper, an ideal omni-directional antenna is assumed at the TX, while a WR-28 waveguide is used at the RX. Let  $G_{i,k}(\phA,\phE,f)$ denote the frequency-dependent far-field beampattern for the $i$-th element in the x direction and $k$-th element in the y direction, $i = 1, 2, \cdots, N_\x$, $k=1, 2, \cdots, N_\y$. Let $\bG(\phA,\phE,f)$ denote the $N_\x \times N_\y$ matrix of beampatterns for the UPA elements. Then the ideal model in (\ref{Hf}) gets replaced with 
\begin{align}
\hspace{-3mm} \bH(f) \! & = \! \! \sum_{n=1}^{N_p} \alpha_n \bG(\phi^\A_n,\phi^\E_n,f) \! \odot \! \left [ \ba_{\Nx}(\theta^\x_{n})  \ba_{\Ny}^{\tdagger}(\theta^\y_{n}) \right ]     e^{-j2\pi \tau_n f} 
\label{Hf_wg_full}
\end{align}
where $\odot$ denotes element-wise product between two matrices of same size. If all antenna beampatterns are identical, as is assumed in this paper, then $G_{i,k}(\phA,\phE,f) = G(\phA,\phE,f)$ for all $(i,k)$, and the non-ideal model in (\ref{Hf_wg_full}) becomes
\begin{align}
\bH(f) & = \sum_{n=1}^{N_p} \alpha_n G(\phi^\A_n,\phi^\E_n,f) \ba_{\Nx}(\theta^\x_{n})  \ba_{\Ny}^{\tdagger}(\theta^\y_{n})   e^{-j2\pi \tau_n f}  
\nonumber \\
& = \sum_{n=1}^{N_p} \alpha_n \bH(f;\mu_n) 
\label{Hf_wg}
\end{align}
where $\mu = (\theta^\x, \theta^\y, \tau)$ (or equivalently, $\mu = (\phi^\A, \phi^\E, \tau))$ and
\begin{align}
\bH(f;\mu) = G(\phi^\A, \phi^\E, f) \ba_{\Nx}(\theta^\x) \ba_{\Ny}^\tdagger(\theta^\y) e^{-j2\pi \tau f}
\label{Hrank1}
\end{align}
is the contribution to $\bH(f)$ defined by $\mu$.

\begin{figure}[htb]
\vspace{-2mm}
\begin{tabular}{cc}
\hspace{-1mm} \includegraphics[width=1.7in]{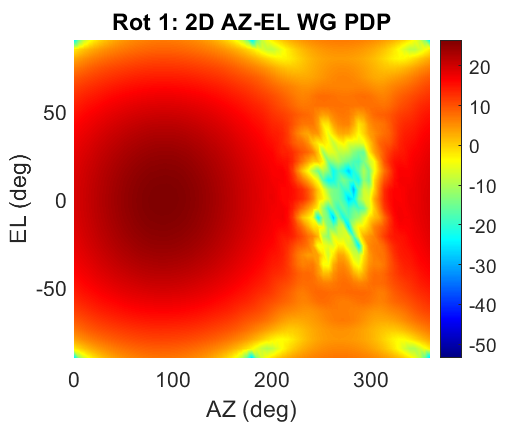} &
\hspace{-7mm} \includegraphics[width=1.7in]{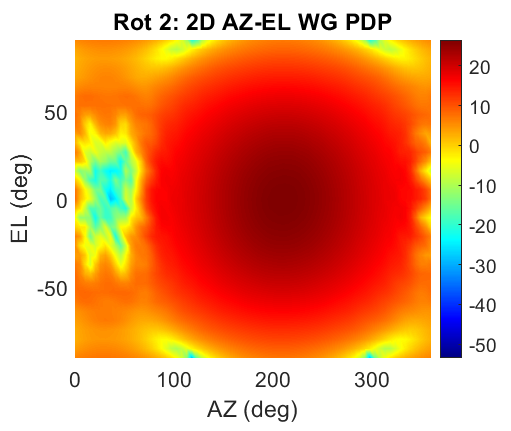} \\
\hspace{-1mm} \includegraphics[width=1.7in]{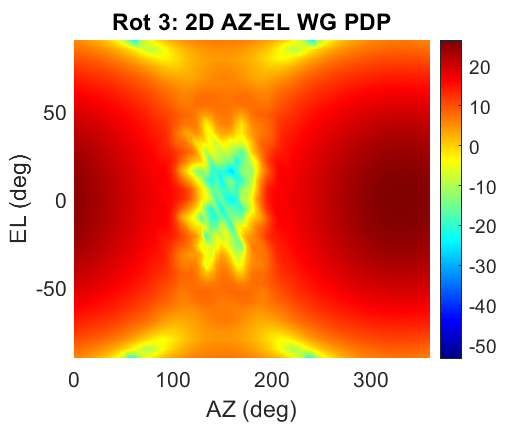} &
\hspace{-7mm} \includegraphics[width=1.7in]{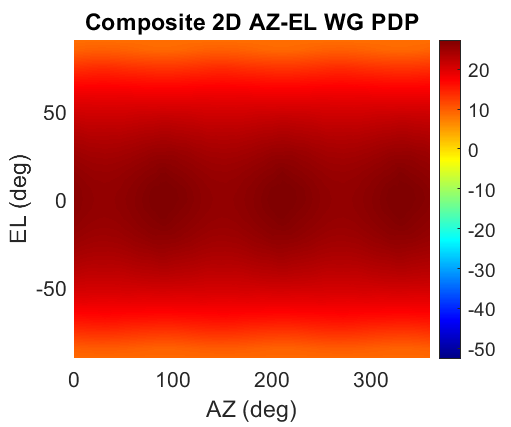} 
\end{tabular}
\vspace{-1mm}
\caption{\footnotesize{\sl The 2D AZ-EL PDPs (dB) of the UPA-element beampattern for the three different UPA orientations, and the composite PDP. Rotation 1 at 90 degrees, Rotation 2 at 210 degrees, and Rotation 3 at 330 degrees.}}
\label{fig:wg_bp_rot}
\vspace{-3mm}
\end{figure}

The measured complex frequency-dependent beampattern $G(\phi^\A,\phi^\E,f)$ for a WR-28 waveguide used in this paper had an initial frequency resolution of 500 MHz and angular resolution of $1^\circ$ in the range $-180^\circ \leq \phi^\A < 180^\circ$, $-90^\circ \leq  \phi^\E \leq 90^\circ$. The beampattern was interpolated to any given $(\phi^\A, \phi^\E)$ using the effective aperture distribution function (EADF) technique \cite{eadf:04}. The frequency samples were interpolated to 10 MHz resolution to match the frequency sampling in the SAMURAI system. The realistic sounder model in (\ref{Hf_wg}) is used for generating the synthetic frequency response samples in Fig.~\ref{fig:method}. 
The 2D power distribution profile (PDP) in AZ-EL of the beampattern is computed as 
\begin{equation}
P_G(\phi^\A, \phi^\E) = \int_{f_c-\frac{W}{2}}^{f_c + \frac{W}{2}} \left |G(\phi^\A,\phi^\E,f) \right |^2 df .
\label{WG_pdp}
\end{equation} 
Fig.~\ref{fig:wg_bp_rot} shows the 2D AZ-EL PDPs of the measured beampatterns for the three UPA orientations as well as the composite beampattern due to all rotations, computed over $W$=1 GHz. 
The full coverage in AZ provided by the three orientations is evident from the composite beampattern. 

\vspace{-3mm}
\section{Estimation of MPCs from Measurements}
\label{sec:extraction}
Consider the non-ideal model (\ref{Hf_wg}) for a static, frequency-selective SIMO channel. The GT MPC data discussed in Sec.~\ref{sec:intro} consists of the physical MPC parameters
\begin{align}
\{\alpha_n, \mu_n  &= (\theta^\x_n, \theta^\y_n, \tau_n) \ : \ n=1, \cdots, N_p\} 
\label{mpc}
\end{align}
which are plugged into (\ref{Hf_wg}) to computationally generate synthetic measurements of $\bH(f)$, which we denote by $\bH_{ms}(f)$.  In practice, the measurements are corrupted by noise
\begin{equation}
\bH_{ms}(f) = \bH(f) + \bW(f)
\label{noisy_meas}
\end{equation}
where the matrix $\bW(f)$ represents thermal noise. The entries of the $\bW(f)$ matrix are statistically independent across the different antenna pairs, and for each entry, $W_{i,k}(f)$, is assumed to be an AWGN (additive white Gaussian noise) process with unit power spectral density, without loss of generality.  The estimation algorithms process $\bH_{ms}(f)$ to generate an estimate of the MPC parameters
\begin{equation}
\{{\hat \alpha}_n, \muh_n = ( {\hat \theta}^{\x}_n, {\hat \theta}^{\y}_n, {\hat \tau}_n) \ : \ n=1, \cdots, {\hat N}_p\}  \  .
\label{mpc_est}
\end{equation}
The sounder makes temporal measurements at the Nyquist rate over  the duration $T$,  resulting in a total of $N =\frac{T}{\tDelta \tau} = {TW}$  samples. A total of $N_\x N_\y$ temporal measurements, each of size $N$, are available for each element of the RX UPA to capture $\bH_{ms}(f)$.  The rest of this section details the algorithms used for MPC estimation from the noisy channel frequency response $\bH_{ms}(f)$ in (\ref{noisy_meas}). The ML formulation of the problem is discussed first, followed by the CLEAN algorithm, least-squares update of MPC amplitudes, and finally the SAGE and RiMAX algorithms. An extension of the system model to fully polarimetric measurements is presented in Sec.~\ref{sec:polarization}.
\vspace{-1mm}
\subsection{Maximum Likelihood Estimation}
\label{sec:ml}
For a given $N_p$, the ML estimate of MPC parameters is given by
\begin{align}
& \{{\hat \alpha}_n, {\hat \tau}_n,{\hat \theta}^{\x}_n,{\hat \theta}^{\y}_n\} 
 = \nonumber  \\
& \arg \!\!\!\! \min_{\alpha_n,\tau_n,\theta^\x_n,\theta^\y_n}  
 \left \|\bH_{ms}(f) -  \sum_{n=1}^{N_p} \alpha_n \bH(f;\mu_n) \right \|^2  \label{mpc_ml}
\end{align}
which operates in a high-dimensional spatio-temporal signal space of dimension $N_o=N N_\x N_\y$  and is computationally prohibitive. It essentially represents a non-linear least-squares optimization problem. The ML estimation in (\ref{mpc_ml}) represents a brute-force search over the $4N_p$ continuous-valued parameters $\{ \alpha_n,\theta^\x_n, \theta^\y_n, \tau_n: n=1, \cdots, N_p\}$ involving the non-linear, non-convex likelihood function on $N_o$-dimensional spatio-temporal vectors. For example, for a sounder equipped  with a UPA at the RX of dimension $N_\x=N_\y=35$, a bandwidth of $W=1$ GHz, and measurement duration of $T=100$ ns, the dimension is $N_o=122500$. Estimation algorithms, such as CLEAN \cite{clean:74}, SAGE \cite{sage:94} and RiMAX \cite{ch_ams:richter:05}, are aimed at taming the computational complexity of ML. It is noted that knowledge of $N_p$ is not assumed by the algorithms in this paper; an estimate of ${N_p}$ is implicitly obtained via the ``stopping criteria'' for the algorithm iterations, as discussed in Sec.~\ref{sec:concepts}.

\vspace{-3mm}
\subsection{CLEAN (Greedy) Estimation: Single MPC update}
\label{sec:greedy}
A simple and often sufficient sub-optimal approach to MPC parameter estimation is the so-called ``greedy'' matching pursuit \cite{mp_mallat:93} which sequentially estimates the dominant MPC components. At the heart of the greedy (similar to CLEAN \cite{clean:74}) approach is the single MPC update: 
\begin{align}
 \{{\hat \alpha}, {\hat \tau},{\hat \theta}^{\x},{\hat \theta}^{\y}\}  & = \arg \!\!\!\! \min_{\alpha,\tau,\theta^\x,\theta^\y}  
 \left \|\bH_{ms}(f) -  \alpha \bH(f;\mu) \right \|^2  \label{ml_greedy}
\end{align}
where $\bH(f;\mu)$ is defined in (\ref{Hrank1}). It is convenient to rewrite (\ref{ml_greedy}) in vector form. Let $\bh = \vc(\bH)$ be the vectorized version of $\bH$ obtained by stacking the columns of $\bH$. Then $\bH(f;\mu)$ becomes
\begin{equation}
\bh(f;\mu) = G(\phi^\A, \phi^\E,f) e^{-j2\pi \tau f} \ba_{\Ny}^*(\theta^\y) \otimes \ba_\Nx(\theta^\x)
\label{hf}
\end{equation}
where $\otimes$ denotes the Kronecker product \cite{Brewer}. By sampling in frequency at $\tDelta f = \frac{1}{T}$, (\ref{hf}) can be expressed as 
\begin{align}
\bh(\mu) & = \left [ \bg_\N(\phi^\A, \phi^\E) \odot \ba_{\N}(\theta^\tau) \right ] \otimes [ \ba_{\Ny}^*(\theta^\y) \otimes \ba_\Nx(\theta^\x)]   \label{bh} \\
\theta^\tau & = \frac{\tau W}{N} \in [0,1)  
\label{th_tau}
\end{align}
where $\bg_\N(\phi^\A, \phi^\E)$ is an $N$-dimensional vector of samples of $G(\phi^\A, \phi^\E,f)$ in frequency, and $\ba_N(\theta^\tau)$  takes the same form as the spatial response vectors in (\ref{st_vec}). The vector $\bh(\mu)$ is an $N_o \times 1$ vector representing the rank-1 channel spatial frequency response with MPC parameter $\mu$. The one-MPC optimization in (\ref{ml_greedy}) can be expressed as 
\begin{align}
 \{{\hat \alpha}, {\hat \mu} \}  & = \arg  \min_{\alpha, \mu}  
 \left \|\bh_{ms} -  \alpha \bh(\mu) \right \|^2  
\label{ml_greedy_vec} 
\end{align}
and the (matched filter) solution is given by
\begin{align}
{\hat \mu}   &= \arg \max_{\mu} f(\bh_{ms},\mu)  \ ; \ f(\bh_{ms},\mu) = \frac{\left | \bh^\tdagger(\mu) \bh_{ms} \right |^2}{\| \bh(\mu) \|^2}  
\label{psi_hat} \\
{\hat \alpha}  & =  \frac{\bh^\tdagger({\muh}) \bh_{ms}}{\| \bh(\muh) \|^2}  \ . \label{alpha_hat} 
\end{align}
The numerator and denominator of $f(\bh_{ms},\mu)$ can be expanded as
\begin{align}
\bh^\tdagger(\mu) \bh_{ms}  = &  \sum_{k=1}^N  \ba_\Nx^\tdagger(\theta^\x) \bH_{ms}(k \tDelta f)\ba_\Ny(\theta^\y) \nonumber \\
 & G^*(\phi^\A,\phi^\E,k \tDelta f) e^{j2\pi \tau k \tDelta f}  \nonumber \\
                           = &  N_\x N_\y   \sum_k H_{b,ms}(\theta^\x, \theta^\y, k \tDelta f) \nonumber \\ 
                           & G^*(\phi^\A,\phi^\E,k \tDelta f) e^{j2\pi \theta^\tau k} \label{hms_h_ip} \\
 \| \bh(\mu) \|^2    = &  N_\x N_\y \|\bg_\N(\phi^\A, \phi^\E)\|^2 \nonumber \\                           
 = & N_\x N_\y   \sum_{k=1}^N \left | G(\phA, \phE, k \tDelta f) \right |^2 
\label{hnorm} 
\end{align}
\begin{align}
& f(\bh_{ms}, \mu)  =  \nonumber \\
  & \frac{ N_\x N_\y \left | \sum_{k} H_{b,ms}(\theta^\x,\theta^\y, k \tDelta f) G^*(\phi^\A,\phi^\E,k \tDelta f) e^{j2 \pi \theta^\tau k} \right |^2}{\sum_k\left |G(\phA, \phE, k\tDelta f) \right |^2 }  \ .
\label{f_bs}                  
\end{align}
That is, the functional $f(\bh_{ms}, \mu)$ in the one-step update is computed in (\ref{f_bs}) by first computing the beam-frequency representation of $\bH_{ms}(f)$ in the spatial dimensions and then multiplying with the conjugate of the element beampattern before computing the beamspace representation in $\tau$.

The CLEAN (greedy matching pursuit) algorithm estimates $K$ MPCs sequentially. 
\vspace{3mm}

\noindent \underline{ALG CLEAN (Greedy):}   
\begin{align}
\mbox{FOR} \ & k=1:K \nonumber  \\
 & \muh_k  =   \arg \max_{\mu} \frac{\left | \bh^\tdagger(\mu) \bh_{ms} \right |^2}{\| \bh(\mu) \|^2}  \ ; \ 
& \alphah_k   =   \frac{\bh^\tdagger({\muh_k}) \bh_{ms}}{\| \bh(\muh_k) \|^2} \nonumber  \\
& \bh_{ms}  \longleftarrow  \bh_{ms}  - \alphah_k  \bh(\muh_k) \nonumber \\
\mbox{END} \ &   \label{alg:greedy} 
\end{align}

Note that  an over-sampled representation of $H_{b,ms}(\theta^\x, \theta^\y, \tau)$ is computed via (\ref{Hb3d}) and (\ref{Hb3d_samp}) to estimate the $K$ dominant MPCs in the greedy algorithm (\ref{alg:greedy}). 

\vspace{-3mm}
\subsection{Least Squares Reconstruction of MPC Amplitudes}
\label{sec:ls_recon}
Once the MPC parameters $\{ (\alphah_k, \muh_k): k =1, \dots, K\}$ for $K$ dominant MPCs have been estimated (using CLEAN, e.g.), a  least squares (LS) update of the MPC complex amplitudes can be obtained to further refine their values.
The measured space-frequency  response is related to the MPC amplitudes  as
\begin{equation}
\bh_{ms} = \bA_{dom} \balpha
\label{ls_mod}
\end{equation}
where $\bh_{ms}$ is the $N_o$-dimensional vector of the measured frequency response,  $\bA_{dom}$ is the $N_o \times K$ matrix whose columns are space-frequency basis vectors corresponding to the estimated MPCs:
\begin{align}
\bA_{dom}  = \left [ \bh(\muh_1), \bh(\muh_2), \cdots, \bh(\muh_K) \right ] \ .
\label{Adom}
\end{align}
The LS estimate for the vector of complex path amplitudes, $\balpha$,  in (\ref{ls_mod}) is given by 
\begin{align}
{\hat \balpha}_{\LS} & = \left ( \bA_{dom}^\tdagger \bA_{dom} \right )^{-1} \bA_{dom}^\tdagger \bh_{ms}  \ .
\label{ls_est}
\end{align}
Note that $\bA_{dom}^\tdagger \bA_{dom}$ is a $K \times K$ matrix and is generally invertible as long as the estimated basis vectors are sufficiently distinct in (\ref{Adom}) and $K < N_o$ which is guaranteed due to multipath sparsity, especially in high-dimensional channels. An LS update improves the estimate of $\balpha$ since the columns of $\bA_{dom}$ are not linearly independent in general; it is also  the best linear unbiased estimator (BLUE) which is efficient from the perspective of the Cramer-Rao lower bound.    

\vspace{-3mm}
\subsection{SAGE Algorithm}
\label{sec:sage}
The greedy part of  the greedy-LS algorithm in (\ref{alg:greedy}) is identical to the CLEAN algorithm - referred to as CLEAN in the following.  The MPC estimates from the CLEAN algorithm are used for initializing  the iterations in SAGE and RiMAX. Let $({\hat \alpha}^{i}_k, {\hat \mu}_k^i)$, $k=1, \cdots, K$, denote the MPC parameter estimates at the $i$-th iteration and ${\hat \alpha}^i_k \bh \left ({\hat \mu}_k^i \right )$ the estimated channel component for the $k$-th MPC at the $i$-th iteration.
A simple path-wise iteration of the SAGE algorithm refines the MPC estimates at the $i$-th iteration via the following two steps ($k=1, \cdots, K$):
\begin{align}
\mbox{E-step:} \ & \bh_{res,k}^i  = \bh_{ms}- \sum_{k' \neq k} {\hat \alpha}^i_{k'} \bh \left ({\hat \mu}_{k'}^i \right )  \label{E_step} \\
\mbox{M-step:} \ &  {\hat \mu}_k^{i+1}  = \arg \max_{\mu} \frac{ \left | \bh^\tdagger(\mu) \bh_{res,k}^i \right |^2}{\| \bh(\mu)\|^2}  
\nonumber \\
 &  
{\hat \alpha}^{i+1}_k =  \frac{\bh^\tdagger \left ({\hat \mu}_k^{i+1} \right ) \bh_{res,k}^i}{ \left \| \bh \left ({\hat \mu}_k^{i+1} \right ) \right \|^2}   \ .\label{M_step}
\end{align}

\vspace{-3mm}
\subsection{RiMAX Algorithm}
\label{sec:rimax}
The RiMAX method offers refinement over CLEAN and SAGE algorithms by using a more general model for the measurements which incorporates diffuse MPCs (DMCs) in addition to the specular MPCs (SMCs) in CLEAN/SAGE:
\begin{align}
\bh_{ms} & =  \bh_{smc} + \bh_{dmc} + \bw  \label{h_rimax}  \\
 \bh_{smc} & = \bA(\bmu)\balpha ; \  \bh_{dmc} \sim \cCN \left ( \bzero, \bR_{dmc} \right ); \  \bw  \sim  \cCN \left (\bzero,  \bI \right ) \nonumber \\
 \bmu  =  \{ & (\theta^\x_k, \theta^\y_k, \tau_k): k = 1,\cdots,K\} ; \ \balpha = [\alpha_1,\cdots,\alpha_{K}]^{\T}   \ . \nonumber
 \end{align}
In the above $\bh_{ms}$, $\bh_{smc}$, $\bh_{dmc}$ and $\bw$ are vectorized versions (sampled frequencies) of the measurement channel matrix $\bH_{ms}(f)$, the SMC channel matrix $\bH_{smc}(f)$, the DMC channel matrix $\bH_{dmc}(f)$, and the noise matrix $\bW(f)$.  The SMC matrix is modeled as deterministic ($\bA(\bmu)$ taking the form (\ref{Adom})) and the DMC matrix is modeled statistically through the covariance matrix $\bR_{dmc}$. The ML estimation problem for (\ref{h_rimax}) is
 \begin{align}
 & ({\hat \bmu}, {\hat \balpha},  {\hat \bR}_{dmc})    = \nonumber \\
  &  
 \arg \min_{\bmu, \balpha, \bR_{dmc}}  \left  [\bh_{ms} - \bh_{smc} \right ]^{\tdagger}  \bR_{dan}^{-1} \left [\bh_{ms}-\bh_{smc} \right ]  + \ln  |\bR_{dan} |   \label{ml_rimax} \\
 & \bR_{dan} = \bR_{dmc} + \bI \nonumber 
 \end{align}
where the subscript $dan$ in $\bR_{dan}$ refers to DMC and noise.
The RiMAX algorithm starts with an initialization of $\bh_{smc}$, using CLEAN or SAGE, as well as $\bR_{dmc}$, and then iteratively updates the estimates of both the SMC ($\bmu$ and $\balpha$) and DMC ($\bR_{dmc}$) parameters:
 \begin{align}
& \mbox{SMC update}: \mbox{fixed} \  {\hat \bR}_{dmc}   \nonumber \\
 & ({\hat \bmu}, {\hat \balpha}) =   
  \nonumber \\
  & 
   \arg \min_{\bmu, \balpha}   [\bh_{ms} - \bh_{smc}(\bmu,\balpha) ]^\tdagger {\hat \bR}_{dan}^{-1}  [ \bh_{ms}-\bh_{smc}(\bmu,\balpha) ] \nonumber \\
& \mbox{DMC update}: \mbox{fixed} \  ({\hat \bmu}, {\hat \balpha})   \nonumber\\
 & {\hat \bR}_{dmc} = 
 \nonumber \\
  &
   \arg \min_{\bR_{dmc}}   [\bh_{ms}-{\hat \bh}_{smc}  ]^\tdagger    \bR_{dan}^{-1}  [\bh_{ms}-{\hat \bh}_{smc} ] + \ln |\bR_{dan}| \ . \nonumber
 \end{align}
An appropriate parametric model for $\bR_{dmc}$ is often used in RiMAX. The RiMAX approach has two advantages. First, over-fitting the data by using too many and possibly weak SMCs is avoided. Second, estimation of $\bR_{dmc}$ also enables more accurate estimation of weaker SMCs. Furthermore, unlike the grid-based search often used in CLEAN and SAGE, the implementation of RiMAX also involves first- and second-order derivatives of the cost function to refine the MPC estimates within a grid cell.

\vspace{-3mm}
\subsection{Polarization Modeling}
\label{sec:polarization}
Thus far, the sounder model and the estimation algorithms have been developed for a single polarization. For a general sounder with dual-polarized antennas, each TX-RX measurement is replaced by a 2 $\times$ 2 system representing the horizontal (H) and vertical (V) polarizations: 
\begin{equation}
\by  = \bH \bx \Longleftrightarrow 
\left [ \begin{array}{c} 
         y^V \\
         y^H \end{array}
         \right ]
         = \left [ \begin{array}{cc}
                  H^{VV} & H^{VH} \\
                  H^{HV} & H^{HH} 
                  \end{array}
                  \right ] 
         \left [ \begin{array}{c} 
         x^V \\
         x^H \end{array}
         \right ] 
\label{polarization}   
 \end{equation}
where $x^V$ and $x^H$ are the signals fed to the V and H ports at the TX antenna, and $y^V$ and $y^H$ are the corresponding signals at the RX antenna. The off-diagonal entries of polarization matrix $\bH$ are the cross-polarization components representing the leakage between H and V components.  Ideally $H^{VH} = H^{HV} = 0$ and in practice $|H^{HV}| \ll |H^{VV}| \ \mbox{and} \ |H^{VH}| \ll |H^{HH}|$. 
The matrix $\bH$ can be further decomposed as 
\begin{align}
\bH & = \left [ \begin{array}{cc}
                  H^{VV} & H^{VH} \\
                  H^{HV} & H^{HH} 
                  \end{array}
                  \right ]  = \bG_R \bA \bG_T 
  \label{Hgen} \\                  
  & 
     =      \left [ \begin{array}{cc}
                  g_R^{VV} & g_R^{VH} \\
                  g_R^{HV} & g_R^{HH} 
                  \end{array}
                  \right ]   
            \left [ \begin{array}{cc}
                  \alpha^{VV} & \alpha^{VH} \\
                  \alpha^{HV} & \alpha^{HH} 
                  \end{array}
                  \right ]                     
             \left [ \begin{array}{cc}
                  g_T^{VV} & g_T^{VH} \\
                  g_T^{HV} & g_T^{HH} 
                  \end{array}
                  \right ]      
\nonumber
 \end{align} 
where $\bG_T$ and $\bG_R$ represent the polarization matrices for the TX and RX antenna and the matrix $\bA$ represents the polarization matrix for the MPC complex PG. The cross-polarization (off diagonal) components  are significantly weaker that the auto-polarization (diagonal) components: 
\begin{align}
g_{\cdot}^{HV} = g_{\cdot}^{VH} &  \ (\mbox{reciprocity}) \ , \ |g_{\cdot}^{HV}| \ll |g_{\cdot}^{VV}|  \ , \ |g_{\cdot}^{VH}| \ll |g_{\cdot}^{HH}| \nonumber \\
|\alpha^{HV}| & \ll |\alpha^{VV}| \  , \ |\alpha^{VH}|  \ll |\alpha^{HH}| \ . \label{xpol}
\end{align} 
To keep things simple, it was assumed that only V component is excited at the TX ($x^H=0$):
\begin{align}
& y^V  = H^{VV} x^V \ , \ y^H = H^{HV} x^V  \label{pol_setup} \\
& \left [ \begin{array}{c}
        H^{VV} \\
        H^{HV}
        \end{array}
       \right ] 
        =  
  \nonumber \\
  &  
 \left [ \begin{array}{cccc}
              g_R^{VV}g_T^{VV} & g_R^{VH}g_T^{VV} &  g_R^{VV}g_T^{HV} & g_R^{VH} g_T^{HV} \\
              g_R^{HV}g_T^{VV} & g_R^{HH}g_T^{VV} &  g_R^{HV}g_T^{HV} &  g_R^{HH} g_T^{HV} 
         \end{array}
   \right ] \hspace{-2mm}
   \left [  
   \begin{array}{c}
        \alpha^{VV} \\
        \alpha^{HV} \\
        \alpha^{VH} \\
        \alpha^{HH} 
   \end{array}
    \right ]  .
    \nonumber        
\end{align}           
Thus, both $y^V$ and $y^H$ are non-zero even when $x^H=0$ and  both need to be processed in general. However, due to computational considerations, only $y^V$ at the RX is utilized by the estimation algorithms in this work. Furthermore, to model an ``omni-polar'' TX antenna, $g_T^{VV} = g_T^{HV} = 1$ was used in the generation of sounder output. 
All four $\alpha$ components via the model $y^{V} = H^{VV} x^V$, where $H^{VV}$ is given in (\ref{pol_setup}), were utilized (and were available as part of the NIST synthetic MPC data described in Sec.~\ref{sec:overview}). Note that $g_R^{VV}$ and $g_R^{HV}$ represent the frequency- and angle-dependent beampatterns and full measured beampatterns for WR-28 waveguide were used in the generation of sounder outputs and algorithms ($g_R^{HV}$ is about 15 dB weaker than $g_R^{VV}$).

\vspace{-3mm}
\section{MPC Estimation from Multiple FoVs}
\label{sec:fov}
In this section, the ML formulation and the CLEAN algorithm are extended for the multiple FoVs. The corresponding extensions for SAGE and RiMAX directly follow. 

\vspace{-3mm}
\subsection{Array Rotations and Beamforming Coordinates}
\label{sec:coords}
\begin{figure}[htb]
\begin{tabular}{cc}
\includegraphics[width=1.4in]{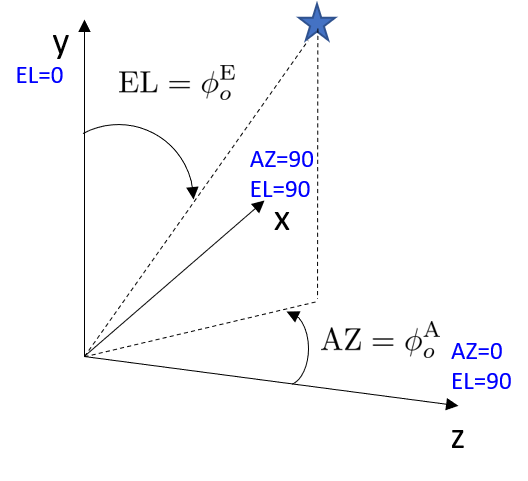}  & 
\includegraphics[width=1.4in]{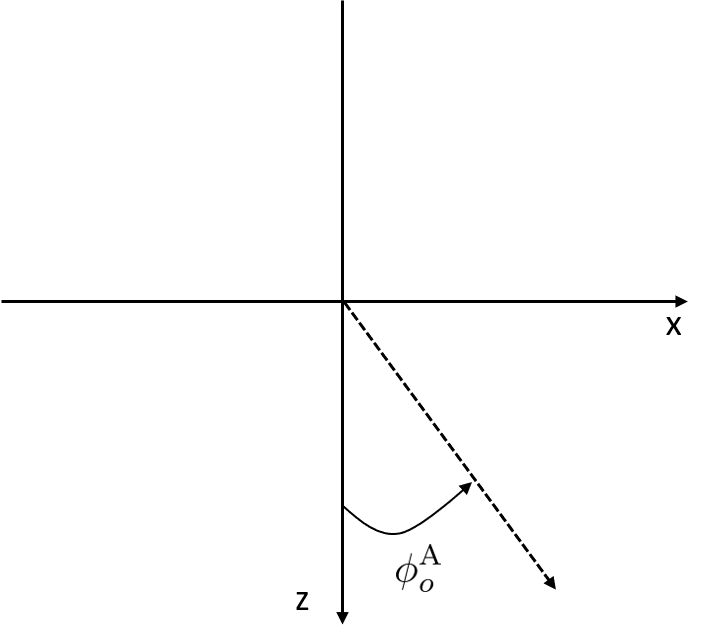} \\
\footnotesize{(a)} & \footnotesize{(b)} 
\end{tabular}
\centerline{\includegraphics[width=2.1in]{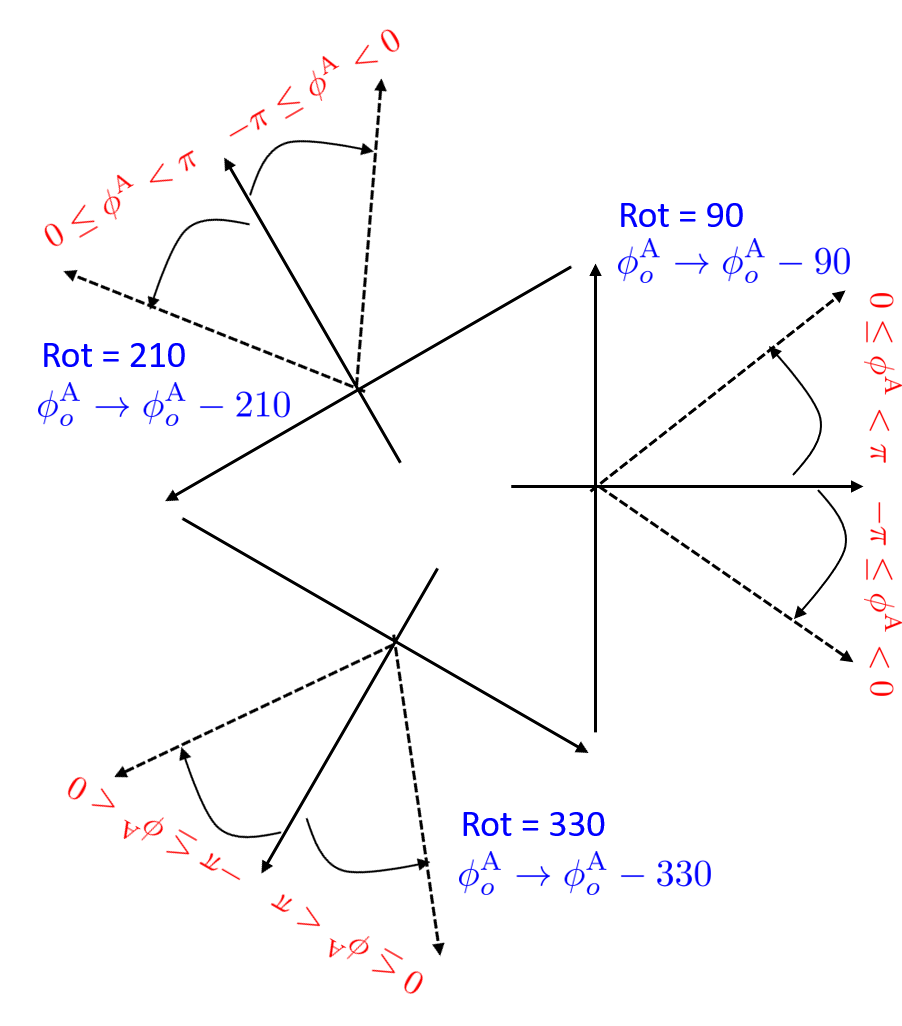}} \\
\centerline{\footnotesize{(c)}}
\caption{\footnotesize{\sl (a) Global AZ-EL coordinates. (b) Global AZ coordinate. (c) Relationship between global AZ coordinates and the local AZ coordinates for the rotated UPAs.}}
\label{fig:azel_coords}
\vspace{-1mm}
\end{figure}

The coordinate system and the global AZ-EL coordinates, $\phi^{\A}_o$ and $\phi^{\E}_o$, are illustrated in Fig.~\ref{fig:azel_coords}.
The global EL angle, $\phi^{\E}_o \in [0,\pi]$, is mapped to the array coordinates $\phi^{\E}$ as  
\[
\phi^{\E} = \frac{\pi}{2} - \phi^{\E}_o \in \left [-\frac{\pi}{2},\frac{\pi}{2} \right ] \ .
\]
The relationship between the global AZ (Fig.~\ref{fig:azel_coords}(b)) and the angles seen by the rotated arrays is illustrated in Fig.~\ref{fig:azel_coords}(c). Let $\phi_{\rot}$ denote the array rotation angle ($90^\circ$, $210^\circ$, $330^\circ$). 
The FoV of the UPA in global AZ coordinates, for a given $\phi_{\rot}$, is
\begin{align}
\fov(\phi_{\rot}) & = \left \{ \phi^\A_o \in \left [\phi_{\rot}-\frac{\pi}{2}, \phi_{\rot}+\frac{\pi}{2} \right ) \right \} \ .
\label{fov}
\end{align}
Specifically, in degrees, $\fov(90) = [0,180)$, $\fov(210) = [120,300)$, and $\fov(330) = [240,420) = [240,60)$. 
For a given global AZ, $\phi^{\A}_o\in [0, 2\pi)$, the local AZ angle seen by the UPA, $\phi^{\A}$, is  
\begin{align}
\phi^{\A} & = \phi^{\A}_o - \phi_{\rot} \in [-\phi_\rot, 2\pi - \phi_\rot) \ .
\label{az_rot}
\end{align}
The rotated AZ coordinates $\phi^{\A}$ get mapped into the local beamforming range $-\pi < \phi^{\A} \leq \pi$ as:
\begin{align}
\phi^{\A} \ & , \ -\pi < \phi^{\A} \leq \pi 
\nonumber \\ 
\phi^{\A} \rightarrow \phi^{\A}  - 2\pi \ & , \ \phi^{\A} > \pi   \\ 
\phi^{\A} \rightarrow \phi^{\A}  + 2 \pi \ & ,  \ \phi^{\A} \leq -\pi  \ .
\label{rot_coords}
\end{align}
For a given array rotation ($\phi_{\rot}$), the estimated AZ angle ${\hat \phi}^{\A} \in [-\pi,\pi)$, is mapped back to the global AZ coordinates as 
\[
{\hat \phi}^{\A}_o = {\hat \phi}^{\A} + \phi_{\rot} \mod (2\pi) \in [0,2\pi)  
\]
and the estimated EL angle ${\hat \phi}^{\E} \in [-\pi/2,\pi/2]$, is mapped back to the global EL coordinates as
\[
{\hat \phi}^{\E}_o = \frac{\pi}{2} - {\hat \phi}^{\E}    \in [0,\pi] \ .
\]

\vspace{-3mm}
\subsection{Multi-FoV Problem Formulation and Single-MPC Update}
\label{sec:fov_prob}
The problem formulation is anchored on a global coordinate system that connects the three views of the rotated UPAs of the same underlying propagation environment. The global MPC coordinates in the 3D angle-delay space $\mu_o = (\phi^{\A}_o, \phi^{\E}, \tau)$  induce the local coordinates for the three arrays: $\mu_i = (\phi^\A_o - \phi_{\rot,i}, \phi^{\E}, \tau)$, $i=1,2,3$ ($90^\circ$, $210^\circ$, $330^\circ$).
The ranges of the global coordinates are $\phi^\A_o \in [0,2\pi)$, $\phE \in [-\pi/2,\pi/2]$, $\tau \in [0,T]$. The only coordinate that changes with UPA rotation is the local AZ coordinate: $\phi^{\A}_i = \phi^\A_o - \phi_{\rot,i}  \ (\mbox{mod} \ 2\pi)$. 

Let $\bh_{ms,i}$ denote the measurement  and $\bh(\bmu_i, \balpha) = \sum_k \alpha_k \bh(\mu_{i,k})$ the corresponding model for the $i$-th rotation where $\bmu_i = (\mu_{i,1}, \mu_{i,2}, \cdots, \mu_{i,K})$, $\balpha = (\alpha_1, \alpha_2, \cdots, \alpha_K)$ and $K$ represents the number of MPCs in the model. Note that $\bmu_i$ are induced by the global harmonic parameters $\bmu_o = (\mu_{o,1}, \mu_{o,2}, \cdots, \mu_{o,K})$, $\mu_{o,k} = (\phi^\A_{o,k}, \phi^\E_{k}, \tau_k)$,  through the rotated AZ angles. For multi-FoV formulation, we stack the measurements and the models into corresponding vectors: 
\begin{align}
\bh_{ms} = \left [ \begin{array}{c} \bh_{ms,1} \\
                                     \bh_{ms,2} \\
                                     \bh_{ms,3} 
                                     \end{array} 
                                     \right ]  \ ; \ 
\bh(\bmu_o,\balpha) = \left [ \begin{array}{c} \bh(\bmu_1, \balpha) \\
                                     \bh(\bmu_2, \balpha) \\
                                     \bh(\bmu_3, \balpha) 
                                     \end{array} 
                                     \right ] \ .
\label{bh_fov}
\end{align} 
The model vector $\bh(\bmu_o,\balpha)$ can be expressed as
\begin{align}
\bh(\bmu_o,\balpha) & = \bH(\bmu_o) \balpha \nonumber \\
\bH(\bmu_o) & = \left [ \bh(\mu_{o,1}), \bh(\mu_{o,2}), \cdots, \bh(\mu_{o,K}) \right ] \nonumber \\
\bh(\mu_{o,k}) & = \left [  \begin{array}{c} 
                          \bh(\mu_{1,k}) \\
                          \bh(\mu_{2,k}) \\
                          \bh(\mu_{3,k}) 
                          \end{array}
                          \right ] \ ; \ k = 1, 2, \cdots, K \ . \label{bh_model}
\end{align} 
The multi-FoV ML estimation problem can be expressed as
\begin{align}
 \hspace{-3mm} \{{\hat \balpha}, {\hat \bmu}_o \}  &\! = \!\arg  \min_{\balpha,\bmu_o}  \sum_{i=1}^3 \left \|\bh_{ms,i} \! - \!  \sum_k \alpha_k \bh(\mu_{i,k}) \right \|^2 
 \nonumber \\
 & 
  = \arg \min_{\balpha,\bmu_o} \left \| \bh_{ms} - \bH(\bmu_o) \balpha \right \|^2  
\label{ml_fov} 
\end{align}
which is of the same form as (\ref{ml_greedy_vec}) and, similar to (\ref{psi_hat}) and (\ref{alpha_hat}), the single-MPC update is
\begin{align}
 \hspace{-3mm} \muh_o   & 
 \! =  \! \arg  \max_{\mu_o}  \frac{\left | \bh^\tdagger (\mu_o) \bh_{ms} \right |^2}{\| \bh(\mu_o)\|^2 }  ; \
{\hat \alpha} \! = \! \frac{\bh^\tdagger({\muh_o}) \bh_{ms}}{\| \bh(\muh_o) \|^2}  \ .
\label{mpc_est_fov}
\end{align}
The LS reconstruction of MPC amplitudes in (\ref{ls_est}) can be extended to the multi-FoV setting by replacing $\bA_{dom}$ with $\bH(\bmu_o)$ and using the multi-FoV definition of $\bh_{ms}$ in (\ref{bh_fov}). The multi-FoV extensions for SAGE and RiMAX, in Secs.~\ref{sec:sage} and \ref{sec:rimax}, follow similarly.

\vspace{-3mm} 
\subsection{Basic Algorithm Design: Key Elements}
\label{sec:concepts}
This section describes the basic algorithm design and the key elements in running the MPC extraction loop. 

\vspace{1mm}
\noindent
{\bf The Basic Algorithmic Loop:}
The basic loop in an MPC extraction algorithm for SMCs is:

\vspace{2mm}

\noindent
{\bf Initialization:} Compute the search regions from  $\bh_{ms}$.
 
\noindent
{\bf WHILE} $k_{mpc} \leq K_{search}$:

\begin{enumerate}

\item
Run the CLEAN algorithm with the single-MPC update within the current search region. Compute the parameters of the candidate MPC: $\muh_{o,k_{mpc}}$ and $\hat{\alpha}_{k_{mpc}}$

\item
{\bf IF} the candidate MPC is valid: 
\begin{itemize} 
\item Do a SAGE update on the MPC parameters $\{ (\alphah_k, \muh_k): k=1, \cdots, k_{mpc} \}$.
\item Do an LS update on the MPC amplitudes $\{\alphah_{k}; k = 1, \cdots, k_{mpc}\}$. 
\item Compute the new residual for $\bh_{ms}$.
\item $k_{mpc} \leftarrow k_{mpc} + 1$ \ .
\end{itemize}
{\bf ELSE IF} the candidate MPC is not valid (one of the rejection criteria is met):
\begin{itemize}
\item Discard the candidate MPC.
\item Go to the next search region. If all the regions in the current set have been exhausted, recompute the regions from the current residual $\bh_{ms}$.  
\end{itemize}
{\bf END (IF)}
\end{enumerate}
\noindent
{\bf END (WHILE)}

\vspace{1mm}

\noindent
Next the key new elements of the basic loop are described: i) region-based searching, and ii) MPC rejection criteria. MPC detection (single-MPC update) and the MPC rejection criteria are the two key elements. Region-based searching can streamline the process. 

\noindent 
{\bf Region-based searching:} The basic idea is to identify disjoint regions in the 3D $\mu$ space based on the ordered (largest to smallest) peaks in the 3D AZ-EL-delay power distribution profile (PDP). Define the following 3D and 1D PDPs:
\begin{align}
P_{\mu_o} (\mu_o) & = P_{\mu_o}(\phi^\A_o, \phE, \tau) =   \frac{\left | \bh^\tdagger (\mu_o) \bh_{ms} \right |^2}{\| \bh(\mu_o)\|^2 } \label{P3d} \\
P_\tau(\tau) & = \int_0^{2\pi} \int_{-\pi/2}^{\pi/2}  P_{\mu_o}(\phi^\A_o, \phE, \tau) d\phE d\phi^\A_o \label{Ptau} 
\end{align}
and the 1D PDPs in AZ and EL,  $P_{\phi^\A_{o}}(\phi^\A_o)$ and $P_{\phE}(\phE)$, are defined similar to the delay PDP in (\ref{Ptau}).
First, the peaks in 1D PDPs of $\phi^\A_o$, $\phE$ and $\tau$ are identified: $\{ \phi^\A_{o,1d,i} \}$, $\{ \phi^\E_{1d,i} \}$, $\{ \tau_{1d,i} \}$. Only peaks that are within a certain threshold $(\gamma_{peak})$ of the largest peak are identified:
\begin{align}
\hspace{-4mm} S_{\tau}(\gamma_{peak}) & = \{\tau_{1d,i}: P_\tau(\tau_{1d,i}) \geq P_{\tau,max} - \gamma_{peak} \}  \label{1d_peaks} 
\end{align}
and $S_{\phi^A_o}(\gamma_{peak})$ and $S_{\phE}(\gamma_{peak})$ are defined similarly.
Now $S_{\mu_o} = S_{\tau} \times S_{\phi^\A_o} \times S_\phE$ defines the set of all potential joint peaks in the 3D space, $\mu_o = (\phi^\A_o, \phE, \tau)$. From that set, a (pruned) set of 3D peaks is identified that are within the threshold $\gamma_{peak}$ of the largest 3D peak: 
\begin{align}
\hspace{-3mm} S_{3d}(\gamma_{peak}) & = \{ \mu \in S_{\mu_o}: P_{\mu_o}(\mu_o) \geq P_{\mu_o,max} - \gamma_{peak} \} \ .
\label{3d_peak_set} 
\end{align} 
The thresholded 3D peaks in $S_{3d}(\gamma_{peak})$ are then sorted in descending order to define the boundaries of the 3D regions: 
\begin{align}
\hspace{-3mm} \mu_{3d,i} & = (\phi^\A_{o,3d,i}, \phi^\E_{3d,i}, \tau_{3d,i}) \ ; \ i = 1, \cdots, |S_{3d}(\gamma_{peak})| \ .
\label{3d_peaks}
\end{align} 
The search regions can be 1D (in delay, e.g.), 2D (in delay-AZ, e.g.) or 3D (in delay-AZ-EL). 

The single-MPC update in the basic loop can  sometimes get stuck at a local maximum due to the grid-nature of the peak search. In such situations, some form of a perturbation is needed to keep the algorithm going. A simple strategy is to move to the next region. 

\noindent
{\bf MPC rejection criteria:}
Sometimes there are good reasons for rejecting a current MPC candidate in the basic loop. After all, the iterative approach is a computationally efficient (and sub-optimal) way of solving the ML estimation problem. There are three main criteria that have been utilized: 

{\bf 1. Power of the candidate MPC:} This occurs when the power of the detected MPC is either below the noise level or its estimated variance violates the Cramer-Rao lower bound.  

{\bf 2. Closeness of the candidate MPC to previously detected MPCs:} If the current MPC candidate's harmonic parameters are too close to those of a previously detected MPC, the resulting 3D channel vectors would be close to linearly dependent and the LS reconstruction may become ill-conditioned. Thus, some measure of closeness is needed to identify such MPCs; e.g., if the candidate MPC's delay, AZ and EL are within half the sounder resolution.

{\bf 3. FoV-violation and peak profile distortion:} An FoV violation occurs when the AZ of a candidate MPC lies within the FoV of a particular array orientation but a different array orientation yields the largest 3D beamformer (BF) output. What this suggests is that another MPC has already been detected near the candidate one and has possibly distorted the 3D shape of the candidate MPC peak. This issue is more likely when the AZ estimate is near the FoV boundaries. One criterion for detecting FoV violation: compare the vector of relative measured BF outputs (for the three orientations) to the vector of expected relative values based on the BF gain. For a given AZ of a candidate MPC, there are three complementary local AZ values for each of the UPA rotations that would yield the same value of the BF outputs. 
Thus, we can compute the relative expected BF powers for the three orientations for each of the three complementary AZ hypotheses. If the measured power vector is closest to the original AZ hypothesis (e.g., using the Kullback-Leibler distance), accept the candidate MPC. Otherwise, reject it.  

{\bf 4. Estimated Relative Variance:} In RiMAX implementation, first-order derivatives of the cost function \eqref{ml_fov} are also computed, from which the deterministic Fisher-Information-Matrix~(FIM) can be obtained. If the estimated value of $\mu$ is close to the global minimum of \eqref{ml_fov}, inversion of the FIM readily provides lower bounds for the variances of the estimated values of the MPC parameters $\mu$, which can, in turn, be transformed to a lower bound for the variance of the absolute value of the corresponding estimated path amplitude \(\vert\hat{\alpha}\vert\). The ratio of the resulting variance to the estimated magnitude \(\vert\hat{\alpha}\vert\), the so-called relative variance, provides a criterion to make a decision on whether to accept or reject the MPC: when the variance is deemed too large compared to the magnitude, as defined by an appropriate threshold, the corresponding MPC is rejected.

\noindent
{\bf Thresholding parameters:} There are two key thresholds that underlie the proposed framework. 
\begin{itemize}
\item
{\bf Peak threshold:} The $\gamma_{peak}$ threshold which determines the lowest peak below the largest peak in defining the regions in a region-based search. This is partly related to the difference between the maximum of a peak and the largest sidelobes. Empirically, $\gamma_{peak} \sim 15-20$~dB.
\item
{\bf Detection threshold:} The $\gamma_{det}$ threshold which defines the weakest MPC amplitude, below the maximum MPC amplitude, that qualifies for a detected MPC.  This threshold depends on the operational SNR  for the MPC: $\min(\snr) \leq \gamma_{det} \leq \max(\snr)$.    
\end{itemize}

\noindent
{\bf Stopping criteria:} Given the iterative nature of all three algorithms, criteria for stopping the iterations are important. The most relevant stopping criterion is based on the reconstruction (normalized) mean-squared error (NMSE) defined in (\ref{error}). For any well-designed iterative algorithm, the NMSE ought to be non-decreasing with the number of MPCs extracted; see Fig.~\ref{fig:bs_images}(d). Thus, an appropriate ``tolerance'' threshold (within $(0,0.1)$, e.g) could be used as a stopping criterion: if the NMSE in successive iterations increases, or decreases less than the threshold, the algorithm stops. Another stopping criterion is when a sufficient number of new potential MPCs are rejected, per a rejection criterion, in succession. Both of these criteria were utilized in the implementations in this paper. Stopping criteria also implicitly define the number of MPCs extracted by the algorithm.

\noindent
{\bf Single-MPC versus multi-MPC update:} We have described the algorithmic framework in the context of the single-MPC update. However, this could also be generalized to a multi-MPC update, albeit at a higher computational complexity.

\vspace{-3mm}
\section{Metrics for Performance Evaluation}
\label{sec:metrics}
In this section, the metrics for evaluation of the MPC extraction algorithms are discussed. The focus is on two types of metrics: one that uses the GT data available in this study (to gain insight), and another that does not rely on GT data and is hence more appropriate in practice. 
\vspace{-1mm}
\subsection{Path Association - Using Ground Truth}
\label{sec:path}
In this section, a procedure is presented for path association (PA) between the estimated and GT MPC parameters to assess algorithm performance. The PA procedure associates $K_{pa} \leq  \min(N_p,K_{dom})$ GT and estimated MPCs that are closest in the AZ-EL-delay space according to an appropriate distance metric.  Let $\cS_{est} = \{ 1, 2, \cdots, K_{dom}\}$ denote the set of indices for the $K_{dom}$ MPCs returned by the algorithm, $\cS_{phy} = \{ 1, 2, \cdots, N_p\}$ denote the set of indices for the physical GT MPCs, and let $\cS_{pa} = \{ 1, 2, \cdots, K_{pa} \}$ denote the set of indices for the estimated MPCs that are associated with the GT MPCs by the PA procedure.  Let $p:  \cS_{pa} \rightarrow \cS_{phy} $ and $q: \cS_{pa} \rightarrow \cS_{est}$ denote index mappings for the GT and estimated MPCs. The objective is to find the mappings $p$ and $q$ that minimize the cost of PA. 

For given $p$ and $q$, the cost function is defined as
\begin{align}
C_{tot} (p,q) & = \sum_{k=1}^{K_{pa}} C(p_k,q_k)    \label{cost} \\
C(p_k,q_k)  & =      C_{\phA,\phE} (p_k, q_k) +  C_{\tau}(p_k,q_k) + C_{\alpha}(p_k,q_k)  \label{cost_sh}
\end{align}
and the individual pairwise costs are defined as 
\begin{align}
C_{\alpha} (p_k, q_\ell) & = \left ( \frac{20\log_{10}(|\alpha_{p_k}|/|{\hat \alpha}_{q_\ell}|) }{\sigma_\alpha} \right )^2
\label{C_pg} \\
C_{\tau}(p_k, q_\ell) & =   \left ( \frac{ \tau_{p_k} -{\hat \tau}_{q_\ell}}{\sigma_\tau} \right )^2 \label{C_tau}  \\
 \hspace{-1mm} C_{\phA,\phE} (p_k, q_\ell) & \hspace{-1mm} = \hspace{-1mm} \left ( \frac{\cos^{-1}\left (S^\T(\phi^\A_{p_k},\phi^\E_{p_k})S({\hat \phi}^\A_{q_\ell}, {\hat \phi}^\E_{q_\ell}) \right )}{\sigma_{\A,\E}} \right )^2 
\label{C_AE}
\end{align}
\begin{align}
\hspace{-3mm} S(\phA, \phE) & = [ \cos(\phA)\sin(\phE), \sin(\phA) \sin(\phE), \cos(\phE)]^\T \label{S_map} 
\end{align}
where $C_{\phA,\phE}$ represents the geodesic distance between the two points on the unit sphere. The normalization factors in the denominator of each of the cost terms, $\sigma_\tau$, $\sigma_{\A,\E}$, and $\sigma_\alpha$, represent the empirical standard deviations for the corresponding numerators in (\ref{C_tau}), (\ref{C_AE}) and (\ref{C_pg}), respectively.
A different cost function based on individual normalized distances in AZ, EL and delay is proposed in \cite{sayeed_mpc:gcom20} and gives similar results.

The Hungarian algorithm (``matchpairs'' function in MATLAB) is used to find the optimal cost-minimizing mappings. The input to the algorithm are all pair-wise costs, $C(p_k,q_\ell); k = 1, \cdots, N_{p} \ , \ \ell = 1, \cdots, K_{dom}$, 
and an adjustable cost value, $C_{um}$, for unmatched pairs. The algorithm returns the mappings for the $K_{pa} $  associated paths: $p_k$, $q_k$, $k=1, \cdots, K_{pa}$. 

\vspace{-3mm}
\subsection{Normalized Mean-Square Error (NMSE) and the Number of Estimated MPCs}
\label{sec:rec_error}
The normalized mean-square error (NMSE) between the original channel measurement and the reconstruction generated from the estimated MPCs provides a performance metric that does not require knowledge of GT; the measured channel serves as a proxy.  The reconstructed estimate for $\bh_{ms}$ is obtained by plugging the MPC estimates $\{ (\alphah_k, \muh_{i,k}), i=1,2,3 ;  k=1, \cdots, K \}$ into (\ref{bh_model}), and the NMSE is given by 
\begin{align}
{\mathrm{NMSE}} & =  \frac{\sum_{i=1}^3 \left \| \bh_{ms,i}  - \sum_{k=1}^{K} \alphah_k \bh(\muh_{i,k}) \right \|^2}{\left \|\bh_{ms} \right \|^2} \ .
\label{error}
\end{align}
The numerator (MSE) can be expressed in a compact form by stacking the measurements for the three rotations:
$\mse  = \left \| \bh_{ms} - \bh_{rec}({\hat \bmu}) \right \|^2$.  The NMSE is expected to be within $(0,1)$.
\begin{figure}[htb]
\centering
\begin{tabular}{cc}
\includegraphics[width=1.7in]{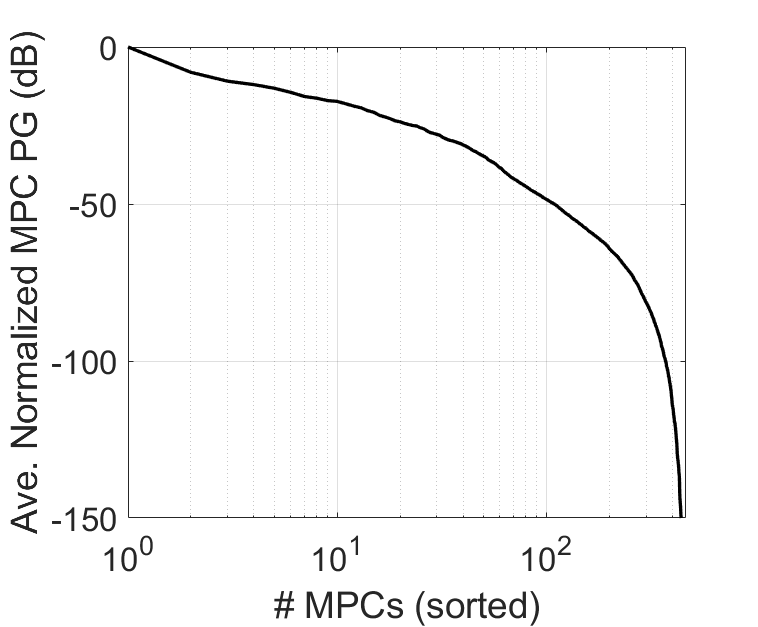} &   
\includegraphics[width=1.7in]{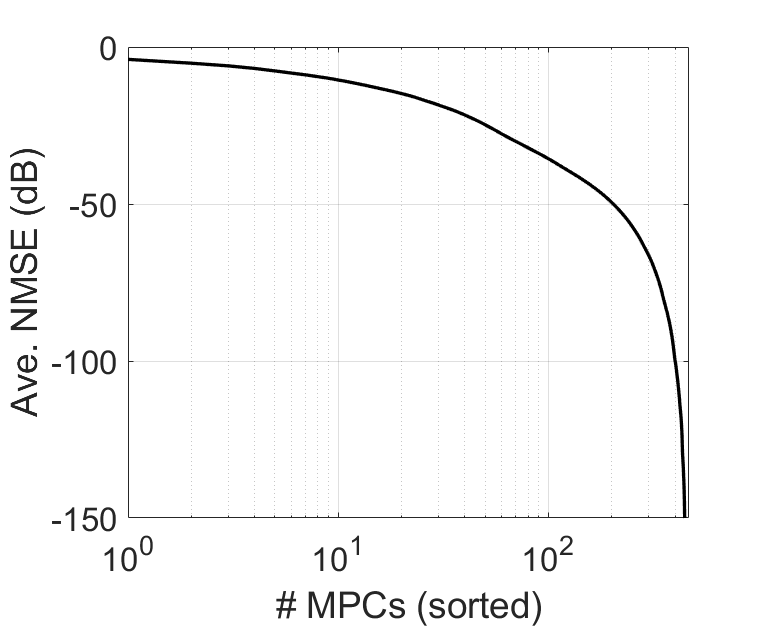} \\[-2mm]
\footnotesize{(a)} & \footnotesize{(b)}
\end{tabular}
\caption{\footnotesize{\sl (a) The average normalized PG, and (b) the average NMSE for a given \# of sorted MPCs.}}
\label{fig:num_mpcs}
\vspace{-3mm}
\end{figure}

The number of estimated MPCs, $K={\hat N}_p$, is another important performance metric as it determines how many salient MPCs have been captured and impacts the NMSE. The GT values of MPC PGs, sorted in decreasing order, are shown in Fig.~\ref{fig:num_mpcs}(a). The value of the theoretical NMSE, for a given number $K$ of sorted GT MPCs captured, is given by 
\begin{equation}
\mbox{NMSE(K)} = \frac{\sum_{k = K+1}^{N_p}|\alpha_k|^2}{\sum_{k=1}^{N_p}|\alpha_k|^2} \ .
\label{nmse_th}
\end{equation}
The NMSE as a function of a given number ($K$) of sorted GT MPCs is shown in Fig.~\ref{fig:num_mpcs}(b). Both quantities are averaged over all 10  measurements. From Fig.~\ref{fig:num_mpcs}(b), the value of the NMSE achieved, on average, by a given number of MPCs, can be calculated: NMSE (dB) = -10.5 (K=10), -14.8 (K=20), -18.5 (K=30), -21.7 (K=40), -24.8 (K=50), -27.7 (K=60), -30 (K=70).  
The value of $K$, coupled with the values of MPC PGs, also provides a measure of the dynamic range of the algorithm; e.g., from  Fig.~\ref{fig:num_mpcs}(a), $K=100$ corresponds to a range of about $50$ dB.

\vspace{-3mm}
\subsection{Computational Complexity}
\label{sec:complexity}
Generally, in terms of the number of multiplications, CLEAN has the lowest computational complexity, SAGE has the next higher complexity, and RiMAX is the most expensive since the underlying model includes DMCs as well. Consider CLEAN and SAGE first since both are based on the SMC model. Let us assume that the same number of MPCs, $K_{smc}={\hat N}_p$, are extracted by both algorithms. The main computation in CLEAN for each iteration is given by (\ref{mpc_est_fov}) which has a complexity of $\cO(N_o^2 N^{os}_{smc})$ where $N_o = NN_\x N_\y$ is the dimension of the signal space and $N^{os}_{smc}$ is an oversampling factor in the AZ-EL-delay space to accurately locate the maximum value.\footnote{In the implementation, first a coarse maximum is calculated using $N^{os}$ of 2 or 3 and then a local search within a 3D resolution bin centered around the coarse maximum is computed with a higher $N^{os}$ of 500 to 1000 corresponding to oversampling by a factor of 8 to 10 in each dimension.} The complexity of computing a critically sampled 3D beamspace representation, the inner product in (\ref{mpc_est_fov}), is $\cO(N_o^2)$. Thus the overall complexity of CLEAN is $\cO(N_o^2 N^{os}_{smc} K_{smc})$ for $K_{smc}$ iterations, the number of MPCs extracted. 

Similarly, the complexity of SAGE, dominated by the M-step in (\ref{M_step}), is $\cO(N_o^2 N^{os}_{smc} K_{smc}^2)$ since for each new MPC all the extracted MPCs are updated through the E-step in (\ref{E_step}) and the M-step in (\ref{M_step}).

The complexity of the RiMAX algorithms is governed by the SMC update and DMC update steps defined in Sec.~\ref{sec:rimax}, in particular the computation of the inverse and the determinant of the DMC covariance matrix $\bR_{dan}$. While $\bR_{dan}$ is $N_o \times N_o$, representing a 3D DMC in AZ-EL-delay, all current implementations only incorporate the model in the delay domain due to complexity considerations, assuming a separable structure: $\bR_{dan} = \bI_{\Nx} \otimes \bI_{\Ny} \otimes \bR_{\tau}$ where the $N \times N$ matrix $\bR_{\tau}$ captures the (parametric) DMC model in the delay domain. Furthermore, $\bR_{\tau}$ has a Toeplitz structure and its inverse as well as the determinant can be efficiently computed using the Cholesky decomposition with $\cO(N^2)$ complexity ($\cO(N^3)$ in the general case) \cite{STEWART:97}. Let $K_{dmc}$ denote the number of DMC parameters (delay modes) extracted by RiMAX and let $N^{os}_{dmc}$ denote the oversampling factor, or number of iterations in a gradient-based approach, to accurately estimate each DMC parameter.

In the SMC update, $\bR_{dan}$ is fixed and involves computing the inverse of $\bR_{dan}$ and $| \bR_{dan}|$ once for the fixed DMC parameters, both of which have $\cO(N^2)$ complexity. Then the quadratic form $[\bh_{ms} - \bh_{smc}(\bmu,\balpha) ]^\tdagger {\hat \bR}_{dan}^{-1}  [ \bh_{ms}-\bh_{smc}(\bmu,\balpha) ]$ needs to be computed for updating the $K_{smc}$ SMC parameters using CLEAN or SAGE, which has a complexity of $\cO(N_o^2 N^{os}_{smc} K_{smc})$ (CLEAN) or $\cO(N_o^2 N^{os}_{smc} K_{smc}^2)$ (SAGE). This update is done for each new DMC parameters extracted, resulting in an overall complexity for the SMC update of $\cO(N_o^2 N^{os}_{smc} K_{smc}K_{dmc})$ (CLEAN) or $\cO(N_o^2 N^{os}_{smc}K_{smc}^2 K_{dmc})$ (SAGE).

The DMC udpate consists of computing the quadratic form involving $\bR_{dan}^{-1}$ as well as the determinant $| \bR_{dan}|$ for $K_{dan}$ DMC parameters (delay modes) extracted. The complexity of computing quadratic forms with $\bR_{dan}^{-1}$, with fixed SMC parameters, is $\cO(N^2 K_{dmc} N^{os}_{dmc}) + \cO(N_o^2 K_{dmc} N^{os}_{dmc})$, where the first term represents  the inverse and the second term represents the quadratic form. The complexity of computing the determinant is $\cO(N^2 K_{dmc} N^{os}_{dmc})$. Thus, the overall complexity of the DMC update is: $\cO(N^2 K_{dmc} N^{os}_{dmc}) + \cO(N_o^2 K_{dmc} N^{os}_{dmc})$, which is dominated by the second term. 

Comparing the complexity of SMC and DMC steps in RiMAX, we conclude that it is dominated by the SMC step since it involves the product of $K_{smc}$ and $K_{dmc}$. Thus, the overall complexity of RiMAX is given by $\cO(N_o^2 N^{os}_{smc} K_{smc}K_{dmc})$ (CLEAN) or $\cO(N_o^2 N^{os}_{smc}K_{smc}^2 K_{dmc})$ (SAGE).

\vspace{-3mm}
\section{Numerical Results}
\label{sec:results}
In this section, numerical results are presented to demonstrate the integrated framework for MPC parameter extraction with CLEAN, SAGE, and RiMAX algorithms developed in Secs.~\ref{sec:extraction}-\ref{sec:fov}, building on the new system (sounder and channel) model developed in Sec.~\ref{sec:nonideal}.  Four different sounder configurations are considered: two configurations defined by bandwidth $W=$ 1 or 2 GHz, and two defined by the size of the UPA $N_\x=N_\y=$17 or 35. The operational center frequency for all configurations is $f_c=28$ GHz. The element spacing is half-wavelength (3.75 mm) at $40$ GHz as in the SAMURAI system, for which a 17 $\times$ 17 UPA is 6 cm $\times$ 6 cm  and a 35 $\times$ 35 UPA is 12.75 cm $\times$ 12.75 cm, with corresponding spatial resolution in the broadside direction given by $\Delta \phi = 9.7^\circ$ and $4.7^\circ$, respectively.  Results on two types of data are reported:  i) synthetic measurements generated by the SAMURAI sounder model from GT MPC data, and ii) real measurements collected by the SAMURAI sounder in a controlled environment.  

\vspace{-3mm}
\subsection{Synthetic Measurements from Ground Truth MPC Data}
\label{sec:syn_meas}
\begin{figure}[htb]
\vspace{-3mm}
\begin{tabular}{c}
\includegraphics[width=3.2in]{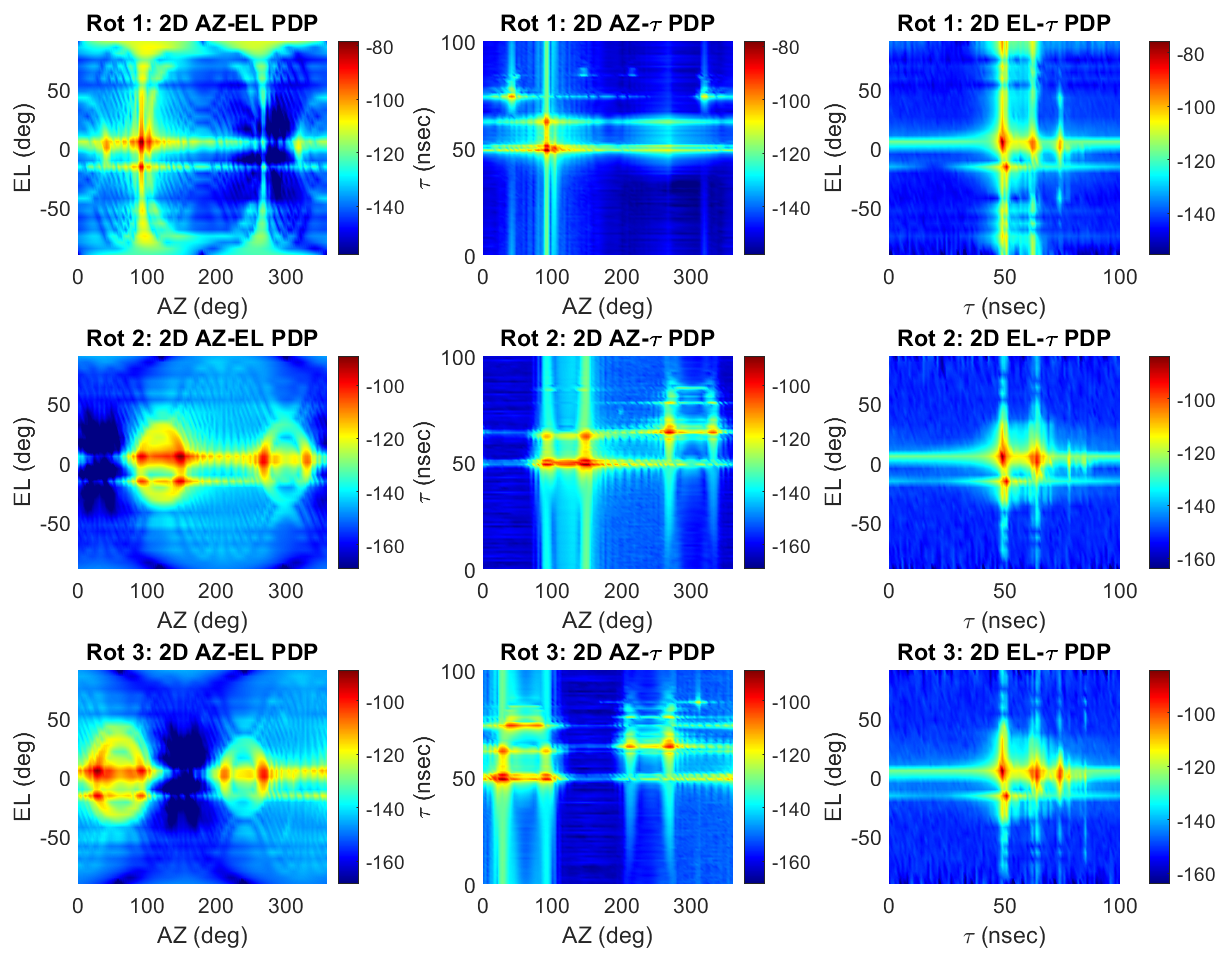} \\[-1mm]
\centerline{\footnotesize{(a)}} \\
\includegraphics[width=3.2in]{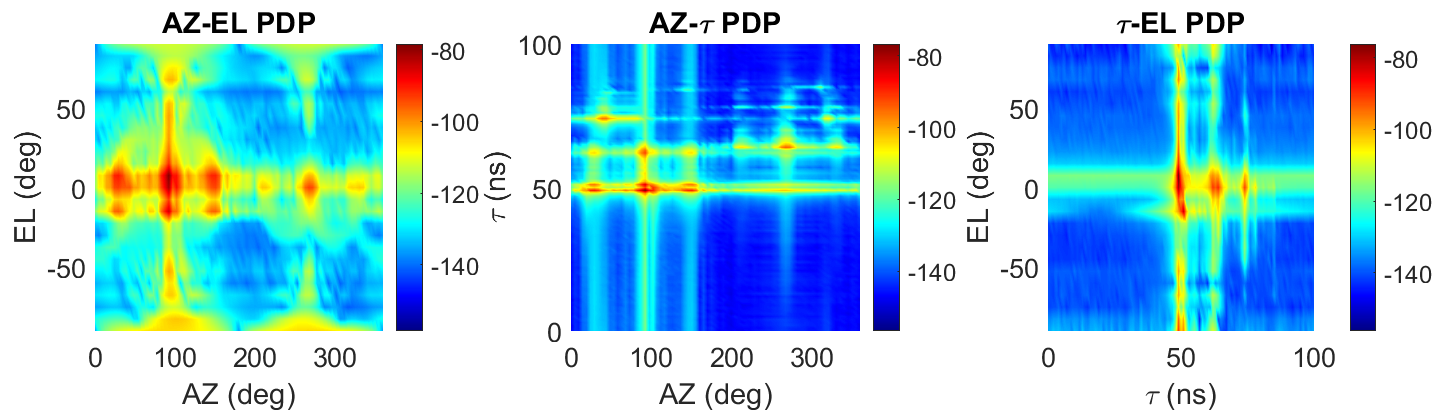} \\[-1mm]
\centerline{\footnotesize{(b)}}\\
\includegraphics[width=3.2in]{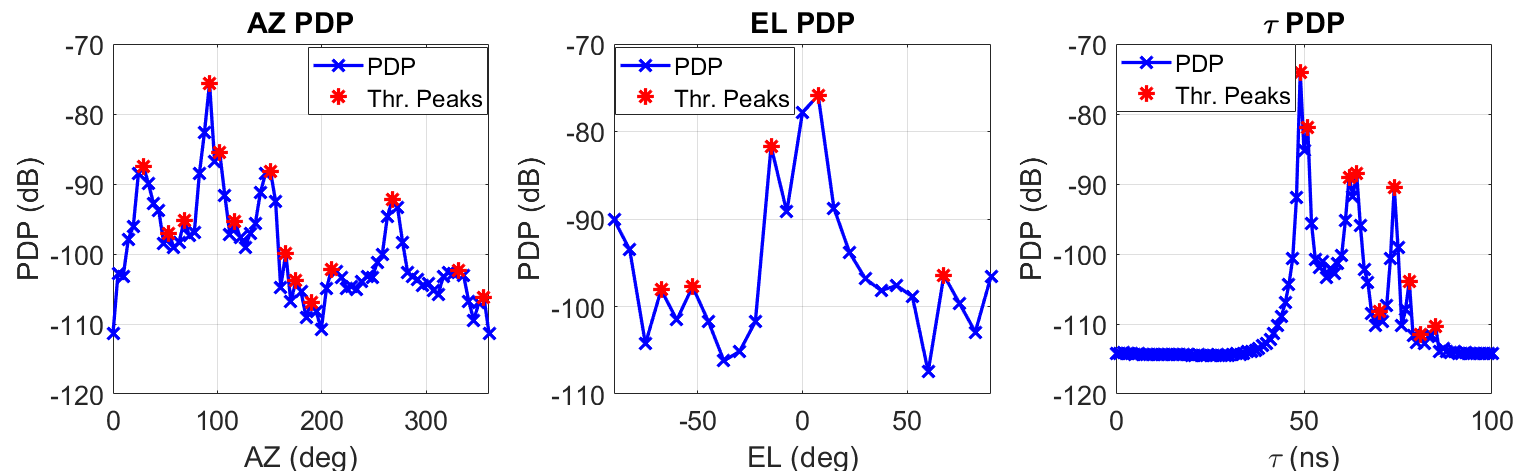} \\[-1mm]
\centerline{\footnotesize{(c)}} \\
\includegraphics[width=3.2in]{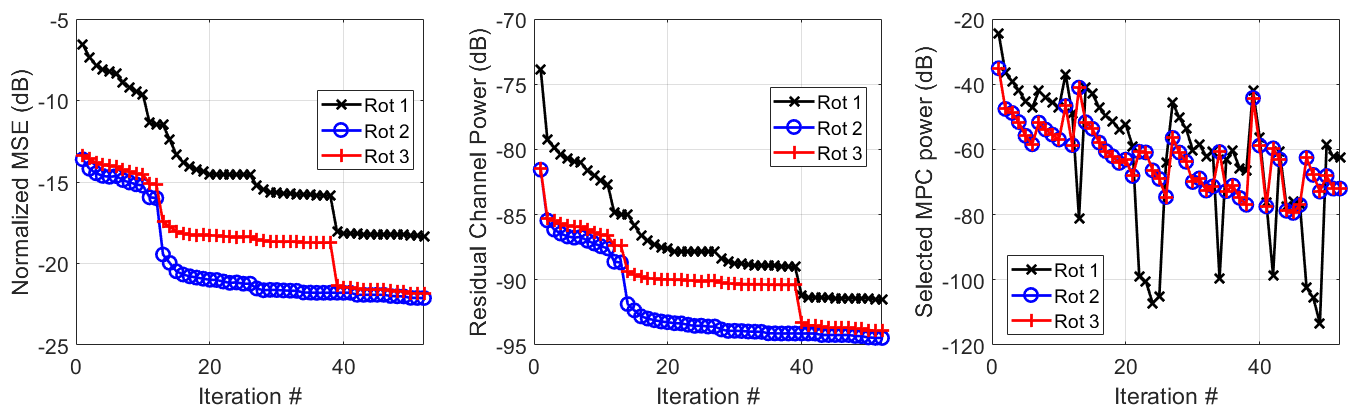} \\[-1mm]
\centerline{\footnotesize{(d)}}
\end{tabular}
\caption{ \footnotesize{\sl (a) The 2D PDPs of the channel for the three UPA orientations. (b) 2D PDPs of the composite channel. (c) 1D PDPs and illustration of peaks for defining the search regions. (d) The NMSE, residual channel power, and the power of the selected MPC  as a function of the iterations for the three UPA orientations.}}
\label{fig:bs_images}
\vspace{-3mm}
\end{figure}
A data set provided by NIST is used for generating the synthetic measurements,  corresponding to a quasi-deterministic channel propagation model reduced from hundreds of measurements in a conference room for 10 different TX-RX locations \cite{nist_mpc:19}; see Sec.~\ref{sec:overview}. The TX is omni-directional and the RX is equipped with three sounder arrays pointing in three distinct directions, centered on disjoint 120-degree sectors in AZ.  Different orientations emphasize different sectors and are jointly processed for estimating the MPC parameters.  A TX power of $30$ dBm is assumed and thermal noise at room temperature, independent across antennas, is added to the measurements.\footnote{The unit dBm is power level expressed in decibels (dB) with reference to one milliwatt (mW): 0 dB = 30 dBm.}

Fig.~\ref{fig:bs_images}(a) shows the 2D PDPs for the three UPA orientations and Fig.~\ref{fig:bs_images}(b) shows the 2D PDPs for the composite metric (all orientations combined) for a particular TX-RX location (scenario 1), $N_\x=N_\y=35$,  $W=1$ GHz. The overall sparsity of the dominant PDP values is quite evident. The sparsity is more accentuated in 2D vs 1D, and most in 3D, as expected. Fig.~\ref{fig:bs_images}(c) shows 1D PDPs for the composite metric and also illustrates the peaks for defining search regions.  Fig.~\ref{fig:bs_images}(d) shows the residual NMSE, residual channel power, and the power of the selected MPC as a function of the iteration index for the CLEAN algorithm for the three UPA orientations. Note that different orientations capture different fractions of channel power, and the NMSE and the residual channel power are non-increasing with the number of iterations, as expected.

\begin{figure}[hbt]
\vspace{-1mm}
\begin{tabular}{cc}
 \includegraphics[width=1.7in]{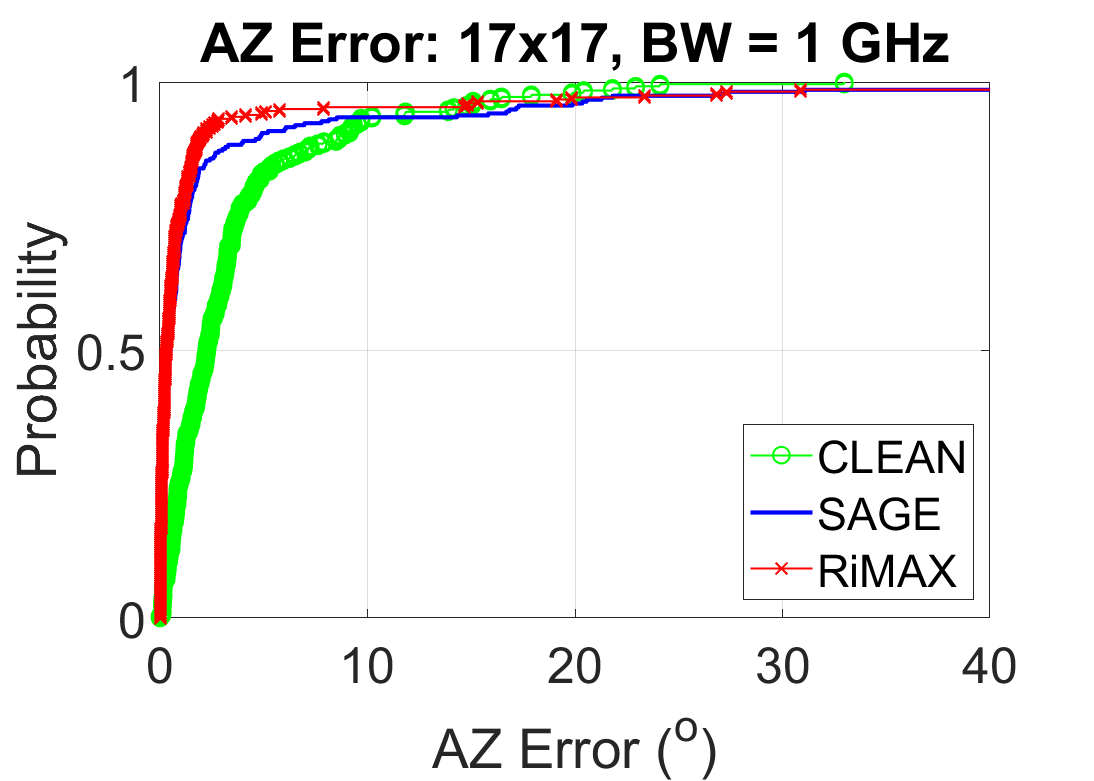} & 
\hspace{-7mm} \includegraphics[width=1.7in]{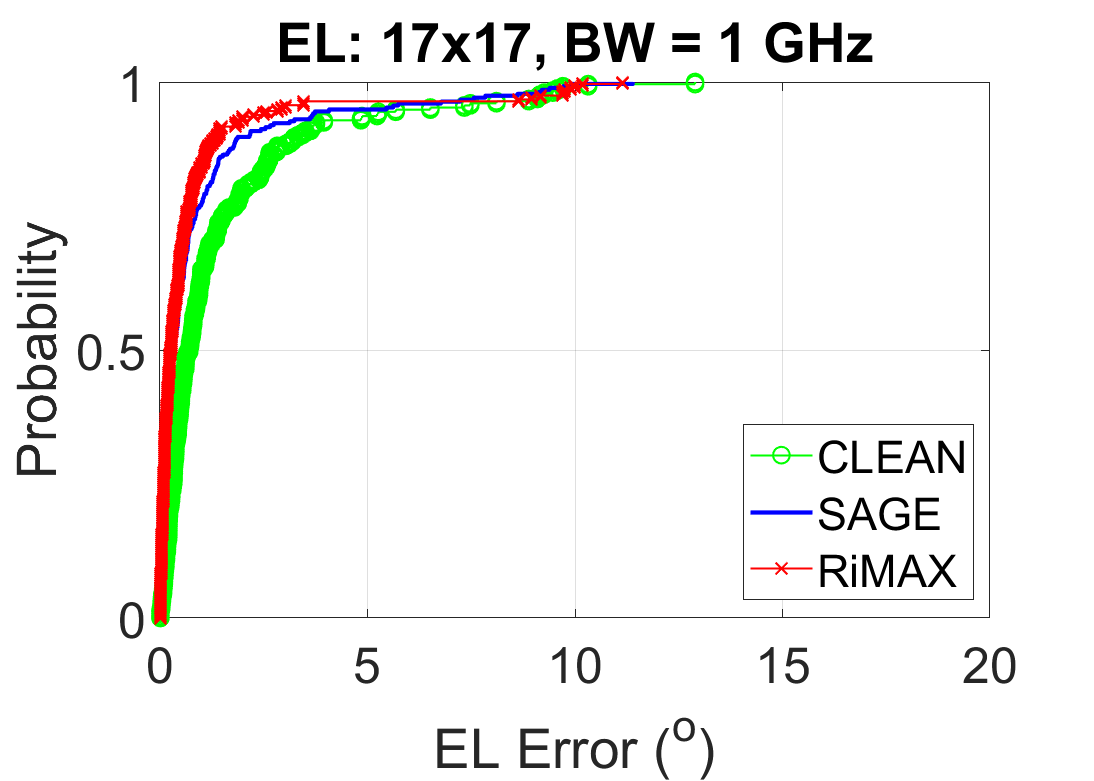} \\ 
 \includegraphics[width=1.7in]{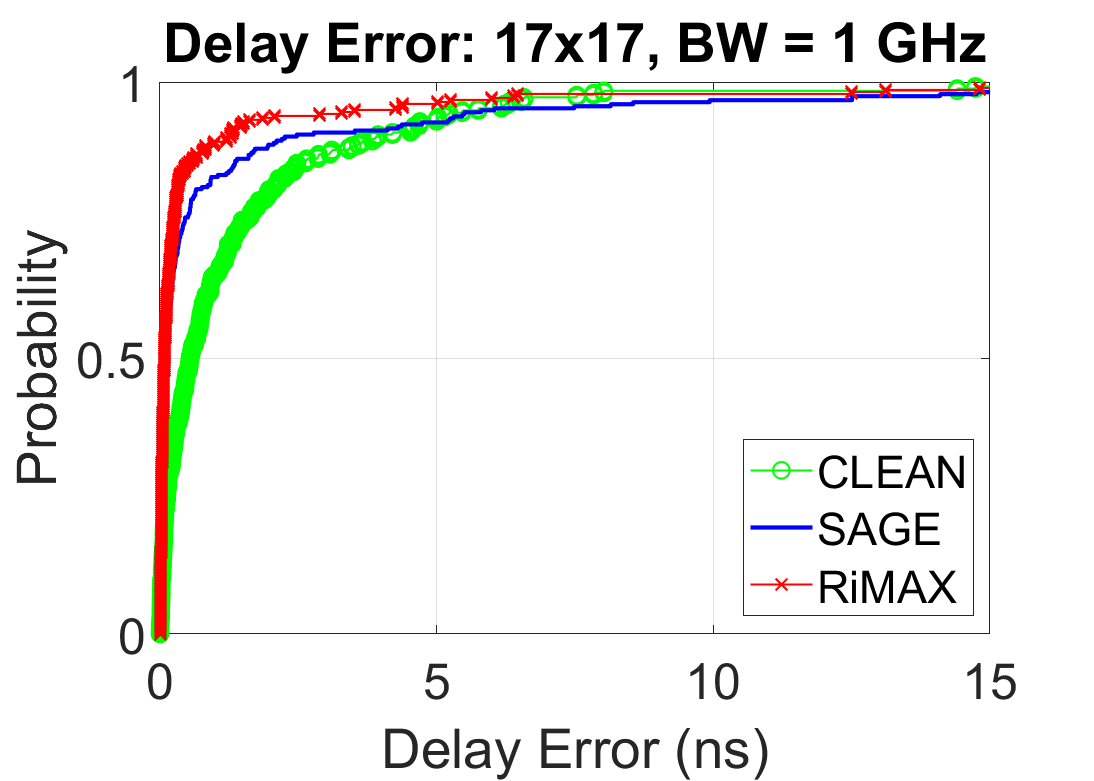} & 
\hspace{-7mm} \includegraphics[width=1.7in]{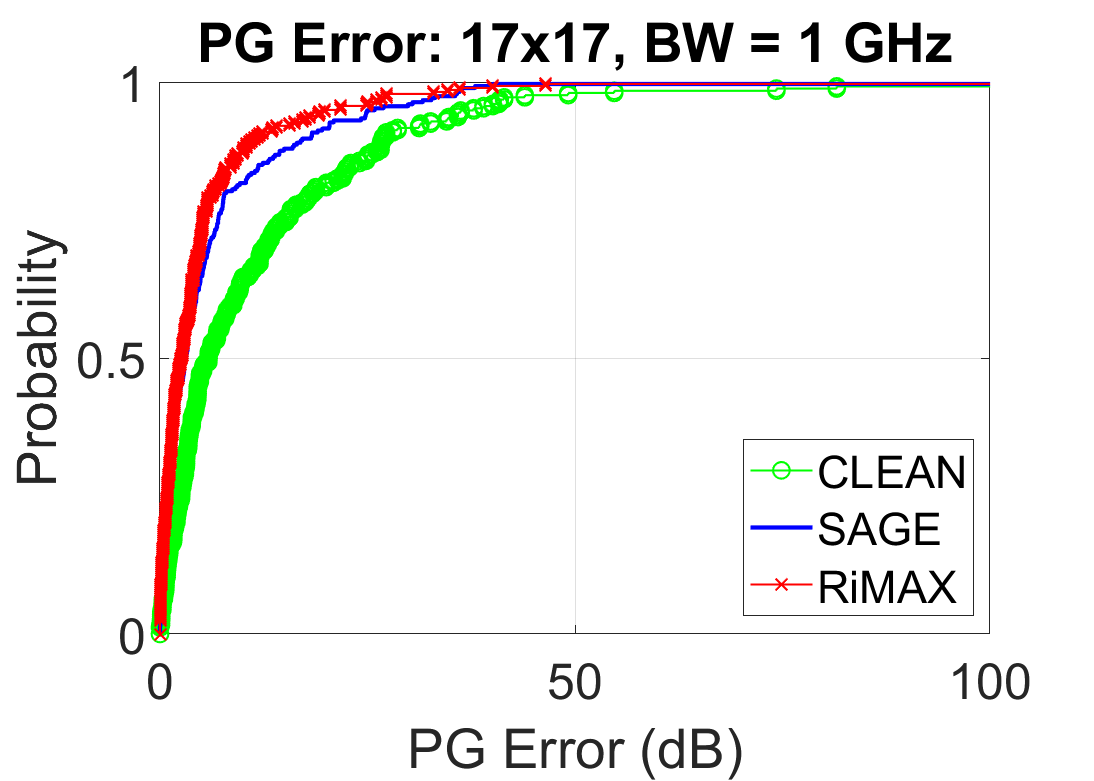}
\vspace{-2mm}
\end{tabular}
\caption{\footnotesize{\sl CDFs of error in AZ, EL, delay and path gain (PG) for the three algorithms for the $17 \times 17$ array and $W=1$ GHz.}}
\label{fig:error_cdfs_17_1G}
\vspace{-1mm}
\end{figure}
\begin{figure}[hbt]
\vspace{-1mm}
\begin{tabular}{cc}
\includegraphics[width=1.7in]{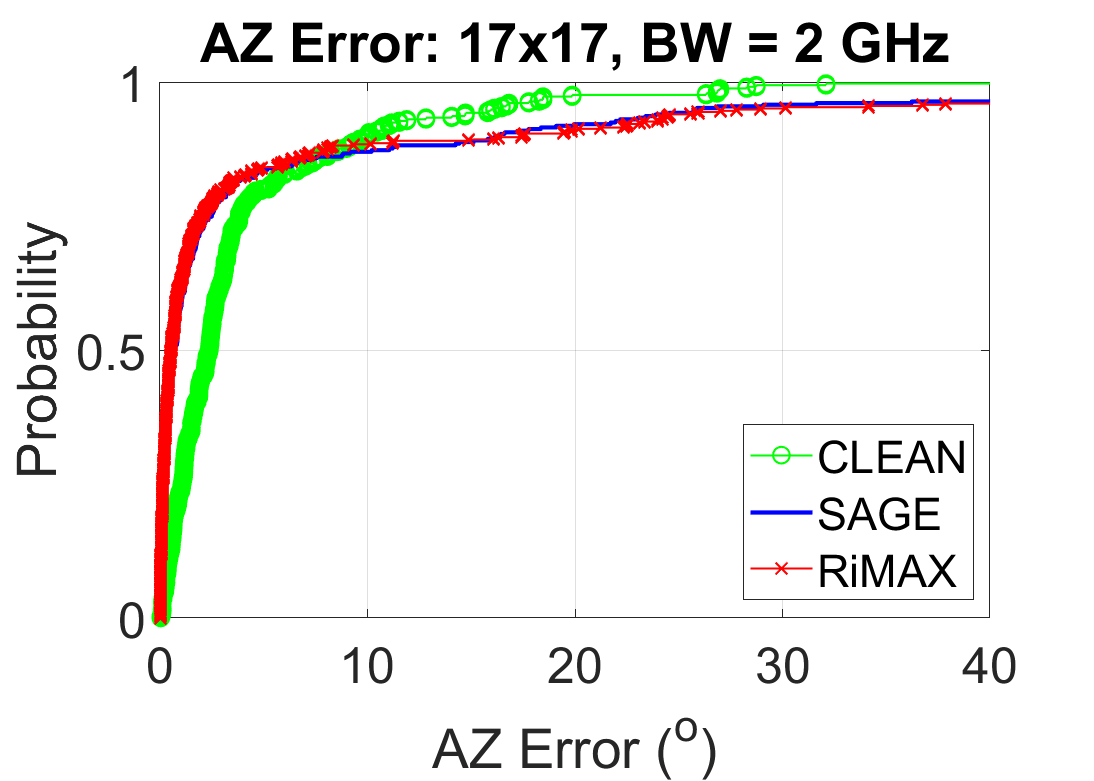} & 
\hspace{-7mm}\includegraphics[width=1.7in]{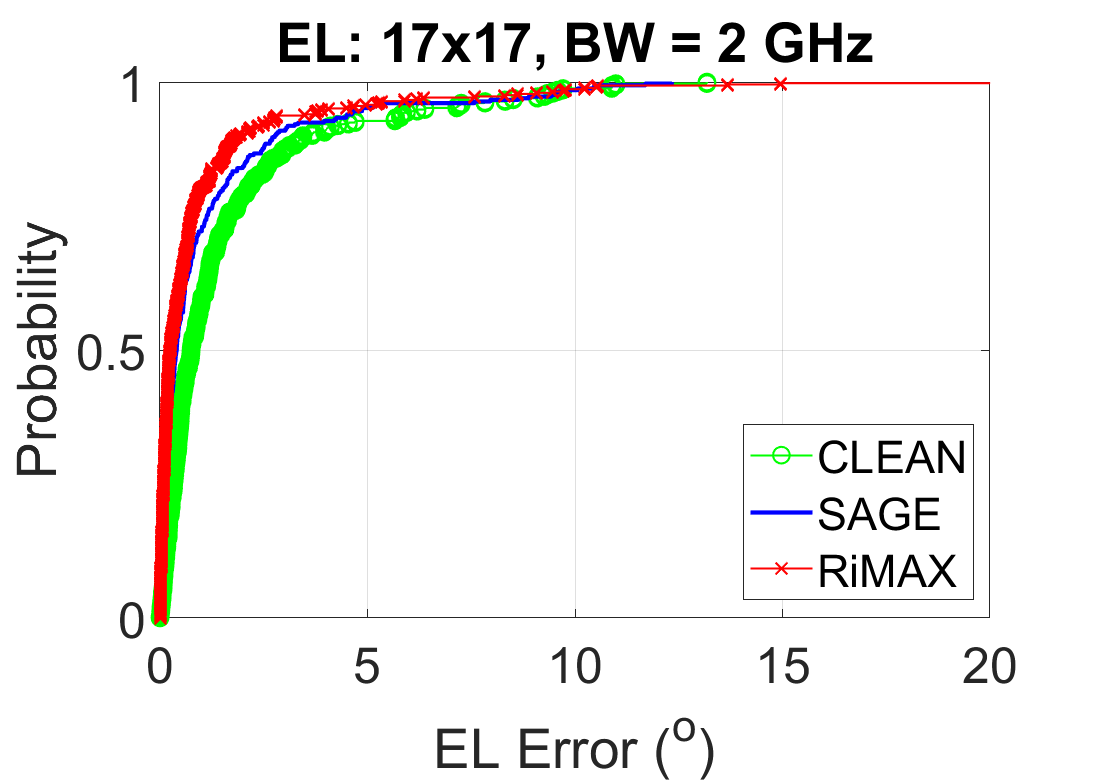} \\
 \includegraphics[width=1.7in]{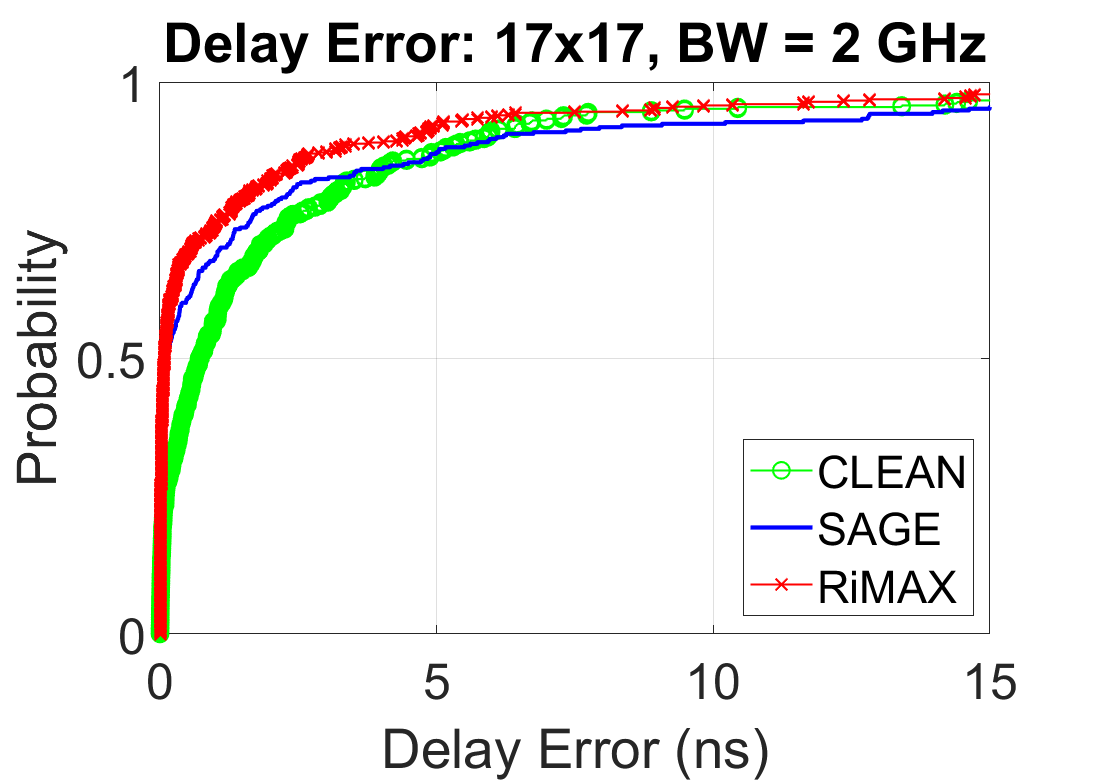} & 
\hspace{-7mm}\includegraphics[width=1.7in]{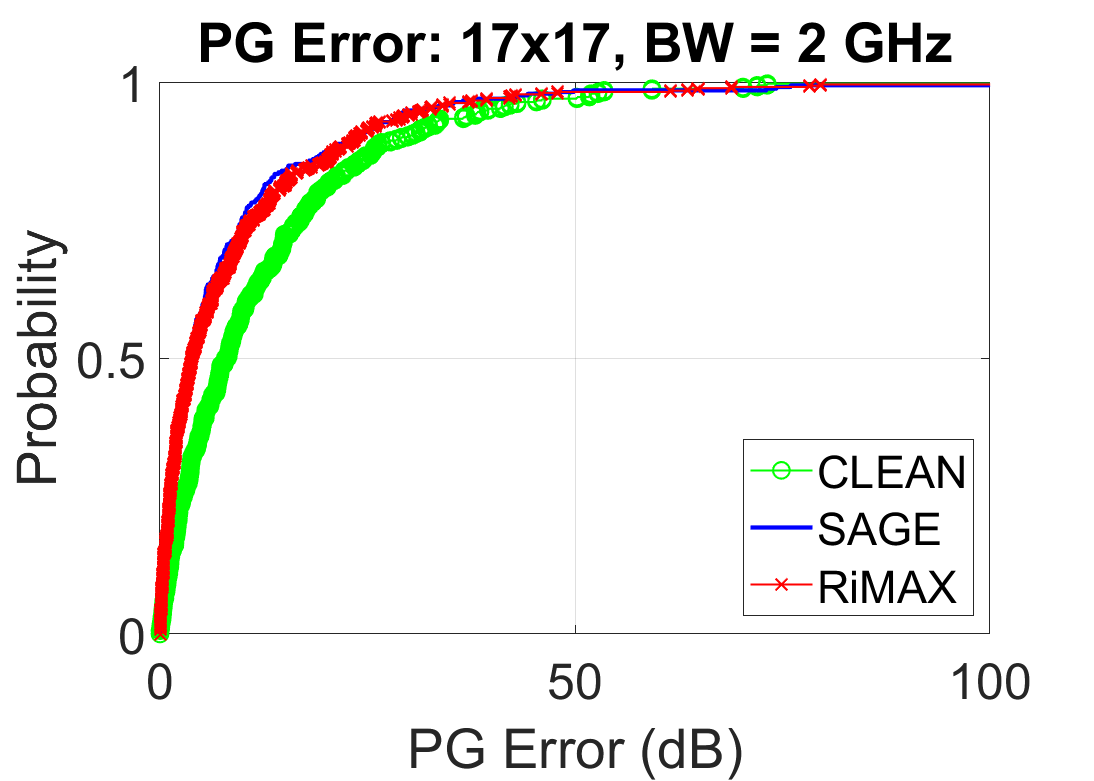}
\vspace{-2mm}
\end{tabular}
\caption{\footnotesize{\sl CDFs of error in AZ, EL, delay and path gain (PG) for the three algorithms for the $17 \times 17$ array and $W=2$ GHz.}}
\label{fig:error_cdfs_17_2G}
\vspace{-1mm}
\end{figure}
\begin{figure}[hbt]
\vspace{-1mm}
\begin{tabular}{cc}
 \includegraphics[width=1.7in]{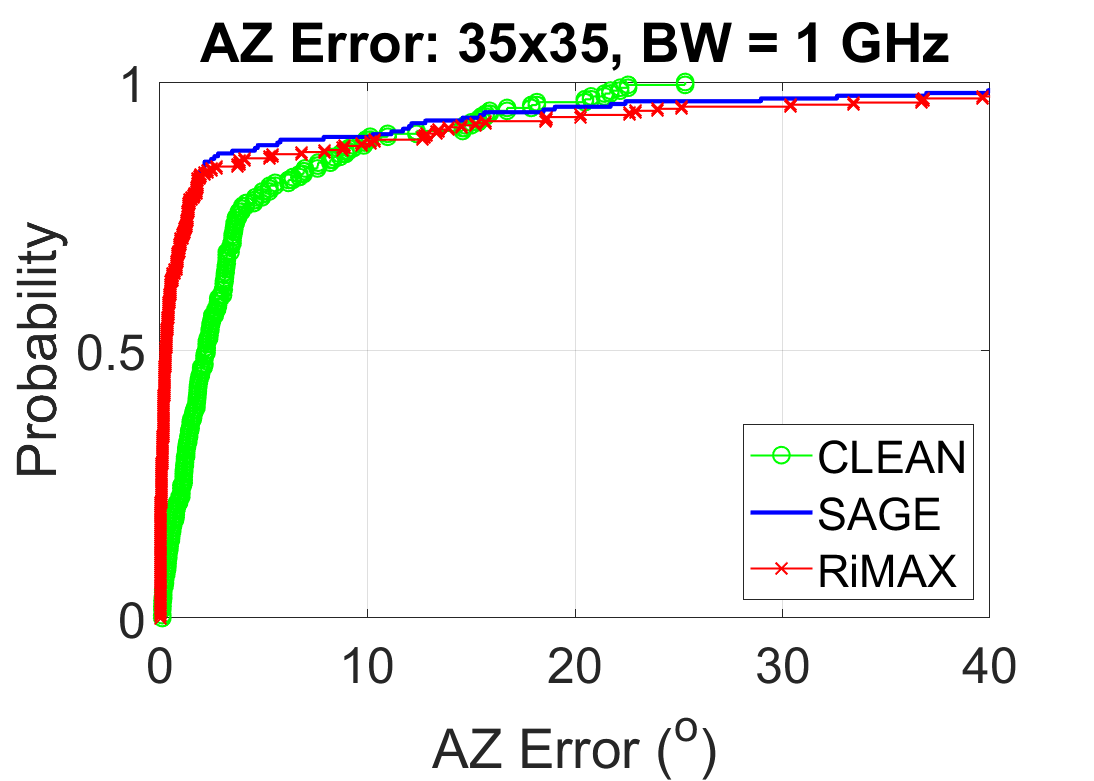} & 
\hspace{-7mm}\includegraphics[width=1.7in]{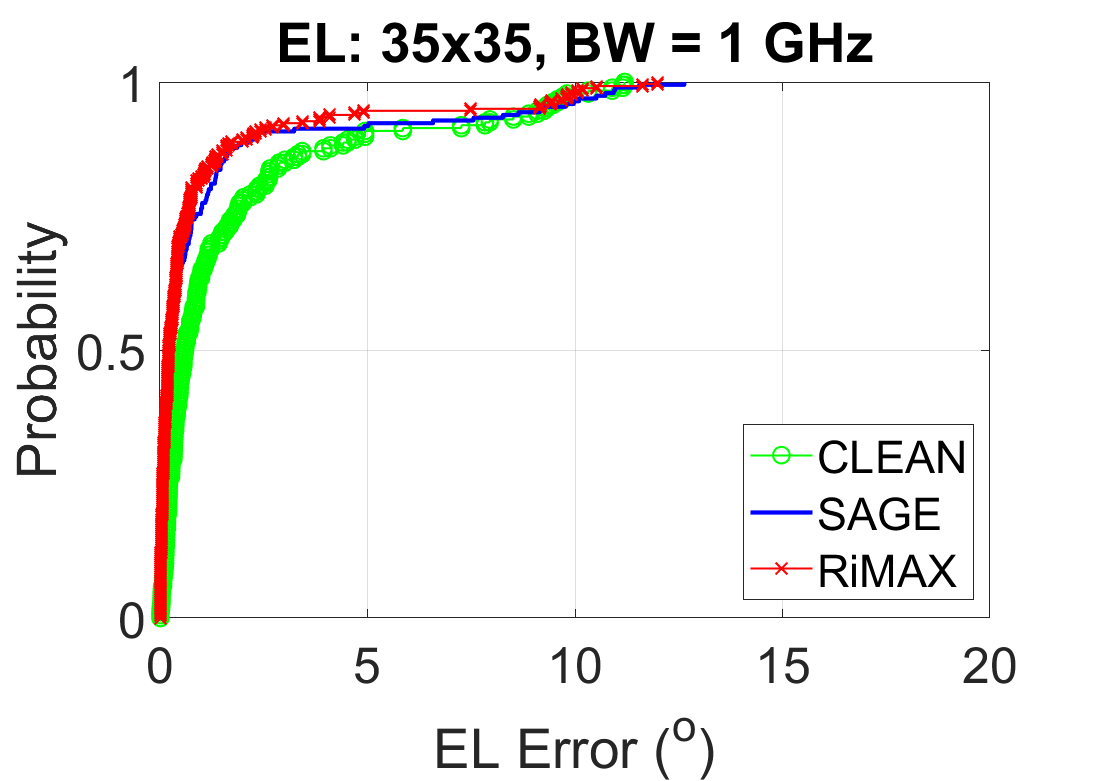} \\
  \includegraphics[width=1.7in]{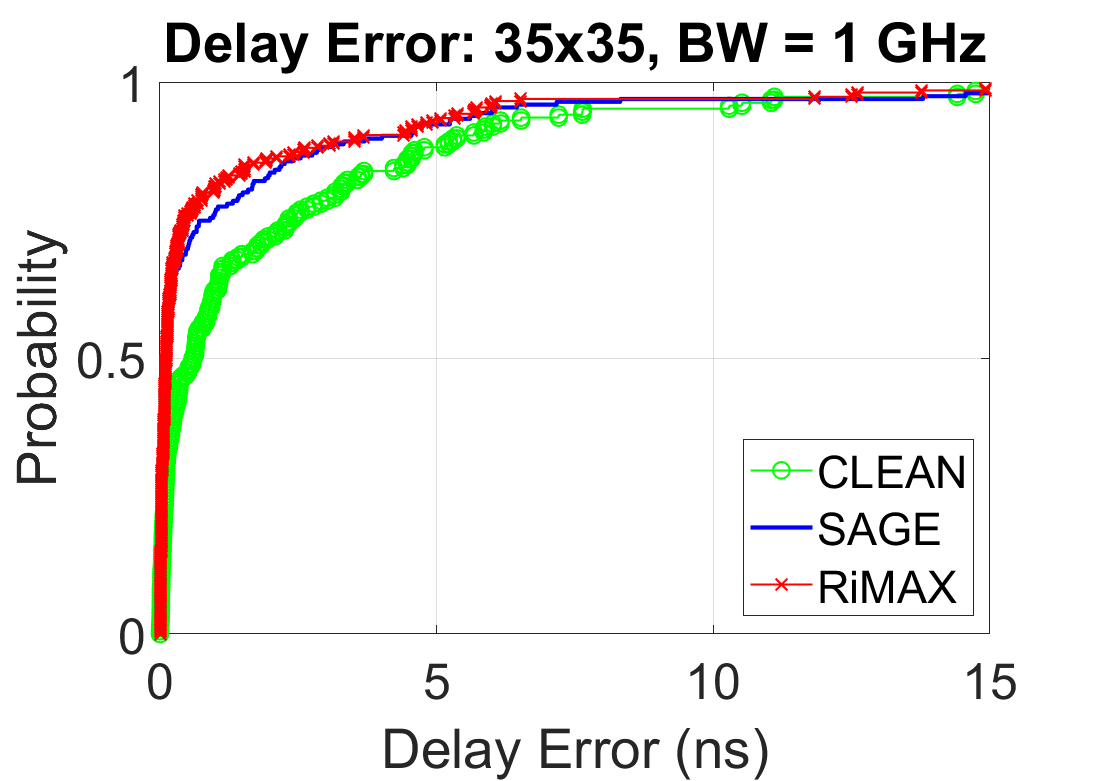} & 
\hspace{-7mm}  \includegraphics[width=1.7in]{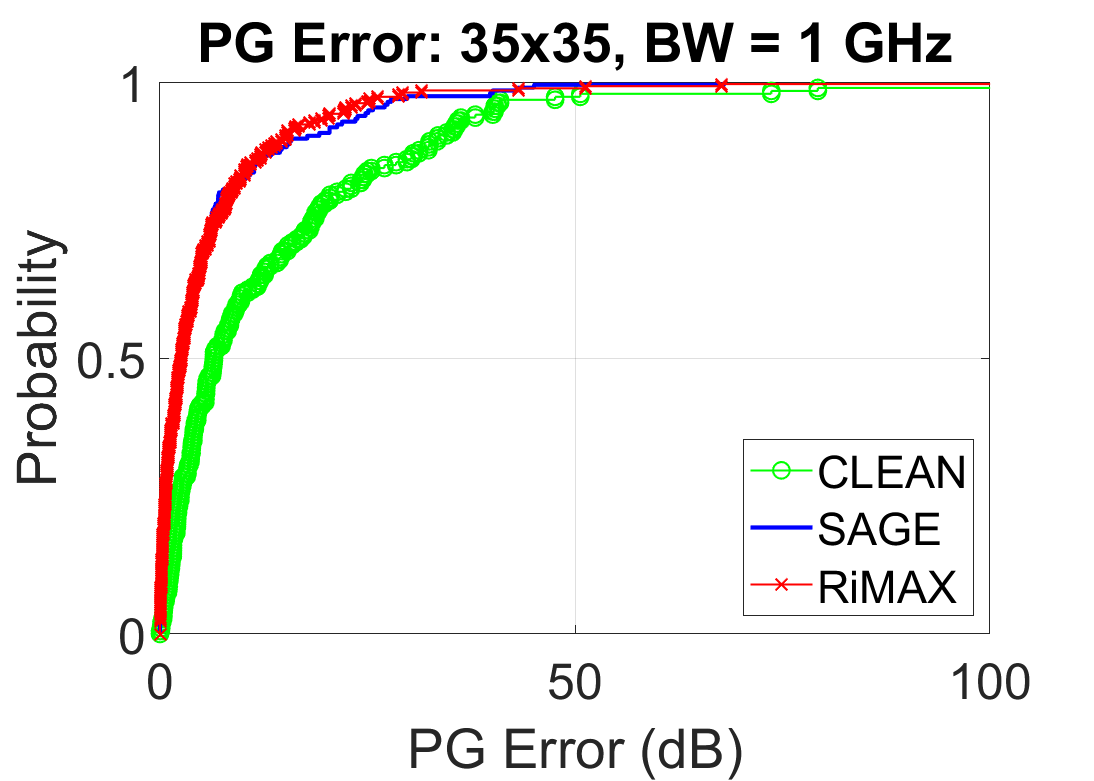}
\vspace{-2mm}
\end{tabular}
\caption{\footnotesize{\sl CDFs of error in AZ, EL, delay and path gain (PG) for the three algorithms for the $35 \times 35$ array and $W=1$ GHz.}}
\label{fig:error_cdfs_35_1G}
\vspace{-1mm}
\end{figure}
\begin{figure}[hbt]
\vspace{-1mm}
\begin{tabular}{cc} 
  \includegraphics[width=1.7in]{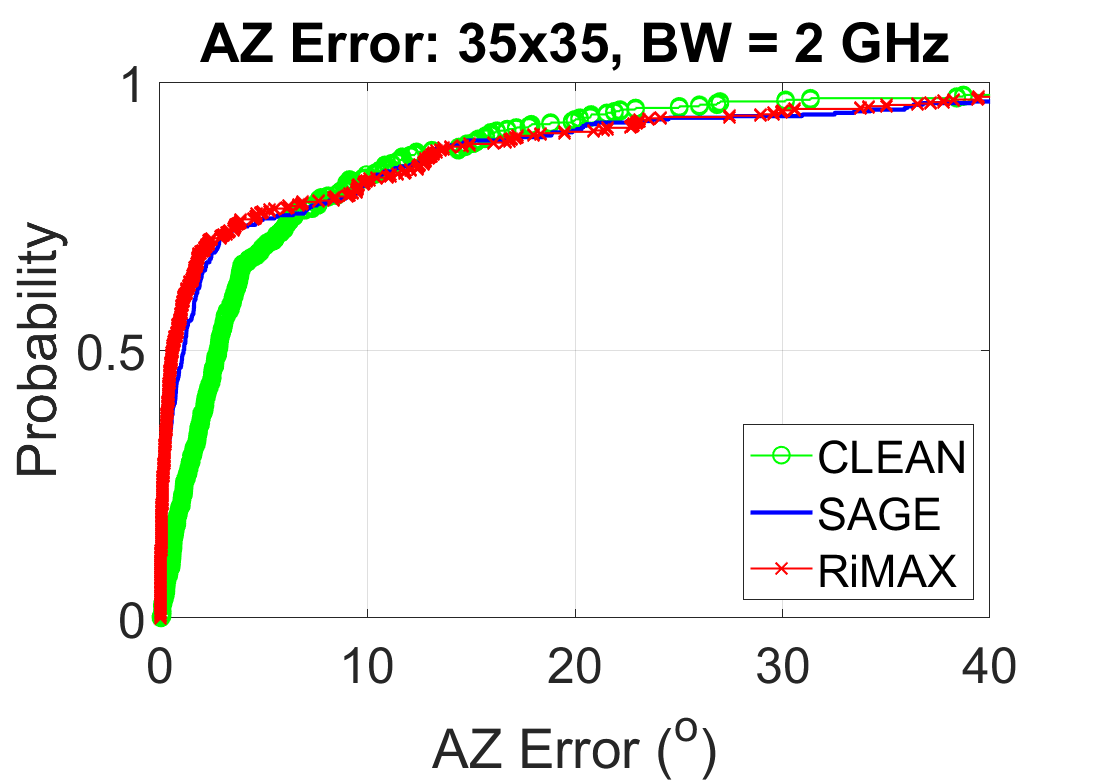} & 
\hspace{-7mm}\includegraphics[width=1.7in]{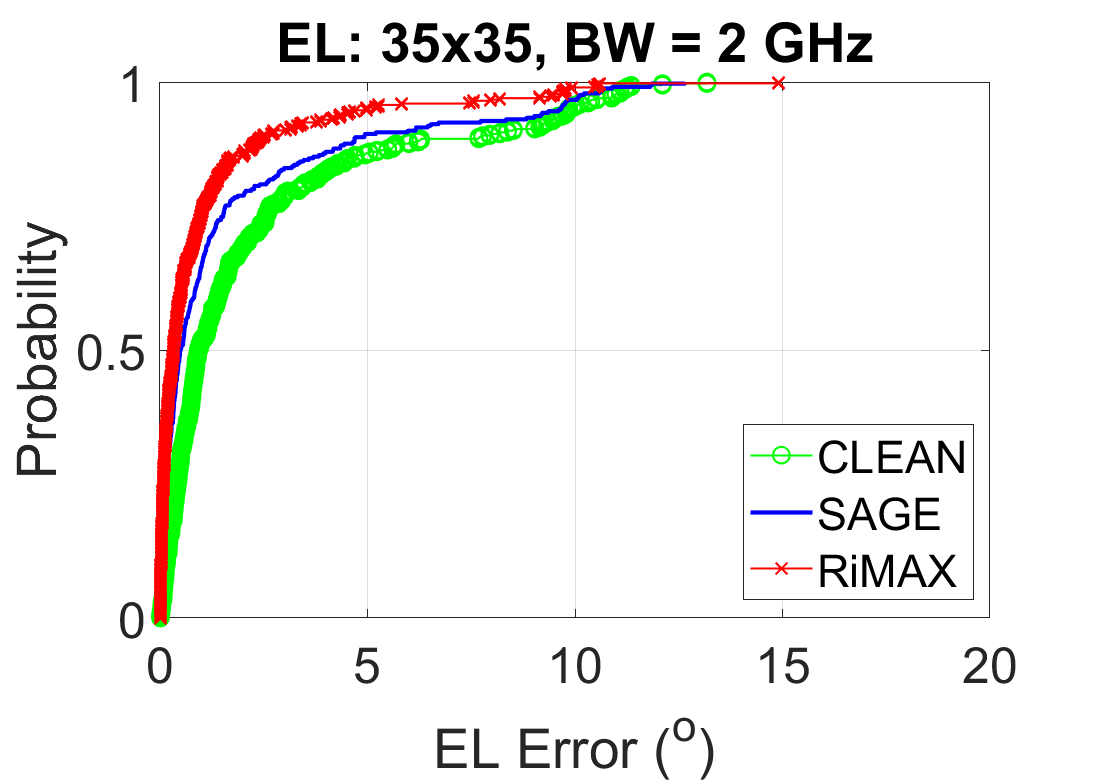} \\
  \includegraphics[width=1.7in]{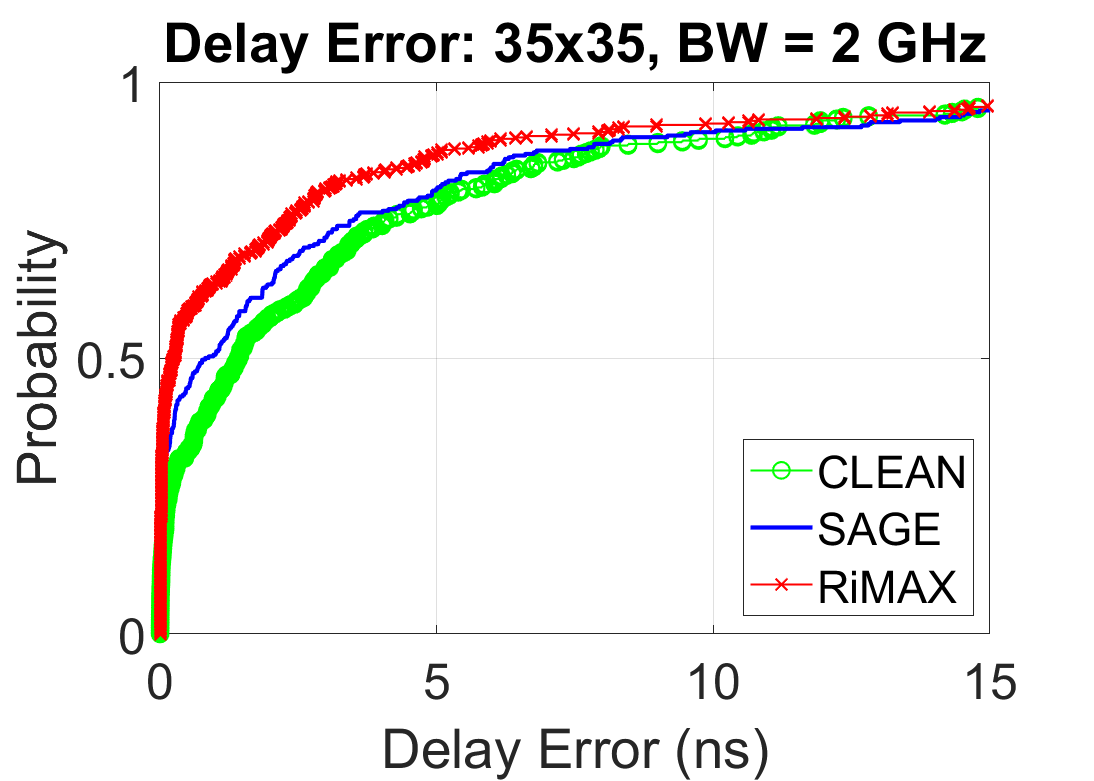} & 
\hspace{-7mm} \includegraphics[width=1.7in]{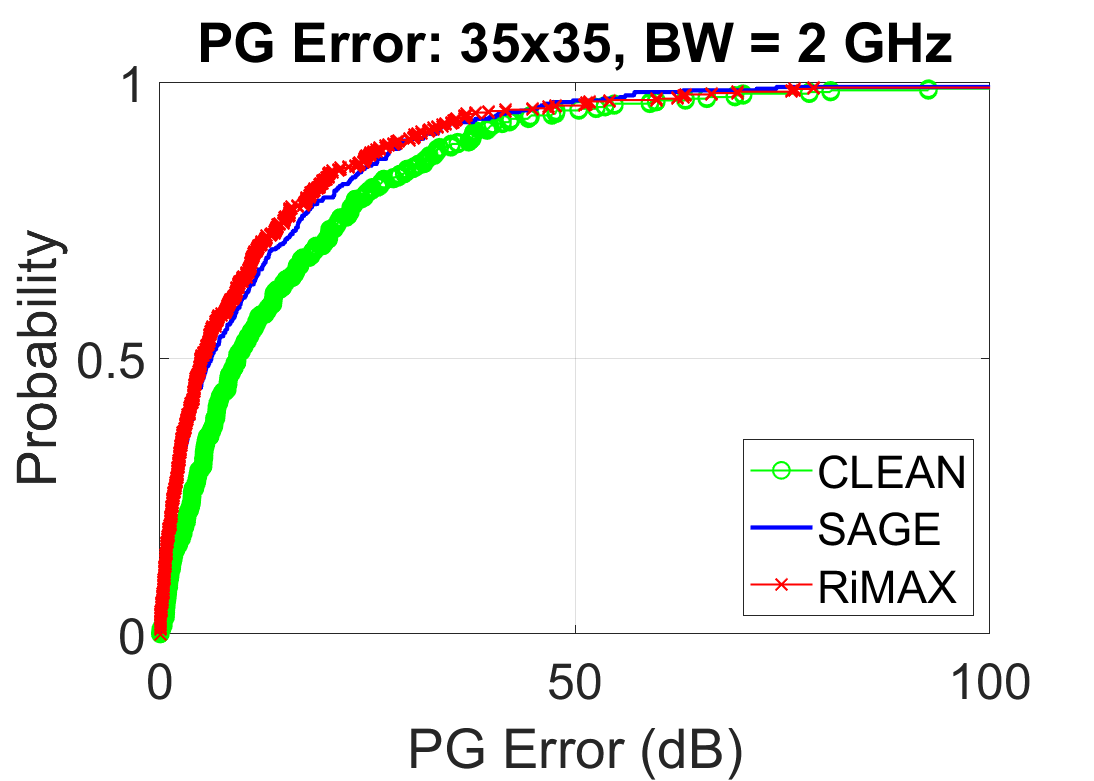}
\vspace{-2mm}
\end{tabular}
\caption{\footnotesize{\sl CDFs of error in AZ, EL, delay and path gain (PG) for the three algorithms for the $35 \times 35$ array and $W=2$ GHz.}}
\label{fig:error_cdfs_35_2G}
\vspace{-1mm}
\end{figure}

The results for the algorithms for the four sounder configurations,  aggregated over measurements for all 10 TX-RX locations, are presented next for the three algorithms: CLEAN, SAGE and RiMAX. Figs.~\ref{fig:error_cdfs_17_1G} and \ref{fig:error_cdfs_17_2G} show the CDFs of the error in the estimated MPC parameters after path association for the 17 $\times$ 17 UPA sounder with $W$=1 GHz and $W$=2 GHz, respectively.  
Figs.~\ref{fig:error_cdfs_35_1G} and \ref{fig:error_cdfs_35_2G} show the corresponding results for the 35 $\times$ 35 UPA.

\begin{figure}[htb]
\vspace{-1mm}
\centering
\begin{tabular}{cc} 
\hspace{-3mm}  \includegraphics[width=1.8in]{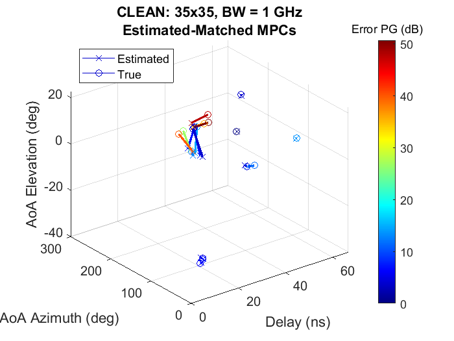} & \hspace{-5mm}
\includegraphics[width=1.8in]{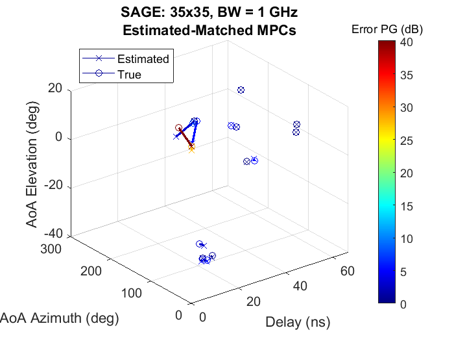} \\[-2mm]
 \footnotesize{CLEAN} & \footnotesize{SAGE} 
 \end{tabular}
\centerline{\includegraphics[width=1.8in]{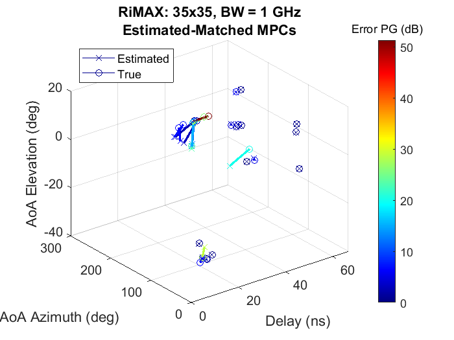}} \\[-2mm]
 \centerline{\footnotesize{RiMAX}}
\caption{\footnotesize{\sl 3D scatter plots of estimated MPC parameters and the corresponding associated GT values for a representative TX-RX location (scenario 1) for the 35 $\times$ 35 array sounder and $W$=1 GHz. The line between the estimated and true MPCs indicates the amount of geometric (AZ, EL, delay) error between the two, whereas their color indicates the size of PG error.}}
\label{fig:error_3d_scatter_35_1G}
\vspace{-1mm}
\end{figure}

Fig.~\ref{fig:error_3d_scatter_35_1G} shows 3D scatter plots of estimated versus GT MPC parameters for the associated paths for the 35 $\times$ 35 array and $W$=1 GHz for a representative TX-RX position. The color coding is based on the error between the PGs of the associated MPCs. Table~\ref{fig:error_table} shows the 50 and 90 percentile errors in AZ, EL, delay and PGs for the four sounder configurations. 

Some general trends from the figures and the table are noted. First, generally the CLEAN algorithm gives quite reasonable estimates of MPC parameters, which are improved upon by SAGE, with further improvements provided by RiMAX. For example,  as seen in Table~\ref{fig:error_table} for the larger array (35 $\times$ 35) and larger bandwidth (2 GHz), CLEAN reports for 50\% of MPCs a maximum angular error of $2.79^\circ$, which is improved by SAGE  to $1.15^\circ$, and RiMAX yields the smallest error of $0.59^\circ$. Similarly, the 50\% delay error for CLEAN is 1.42 ns, which is improved to 0.85 ns by SAGE, and RiMAX yields the smallest error of 0.24 ns. Second, the 50\% errors are generally well-within the sampling resolutions (see (\ref{samp_res})) for the sounder array, whereas the 90\% errors are on the order of the sampling resolutions. Relative to the sampling resolutions, the errors are larger for the larger UPA size and the larger bandwidth. This is partly due to beamsquint effects for wideband arrays that become larger with the product of the fractional bandwidth and array dimension \cite{brady:icc15}. The 17 $\times$ 17 array with a 1 GHz bandwidth is the least prone to beamsquint effects as also evident from errors for the four sounder configurations. 
\begin{table}[hbt]
\vspace{-1mm}
\centerline{\includegraphics[width=3.5in]{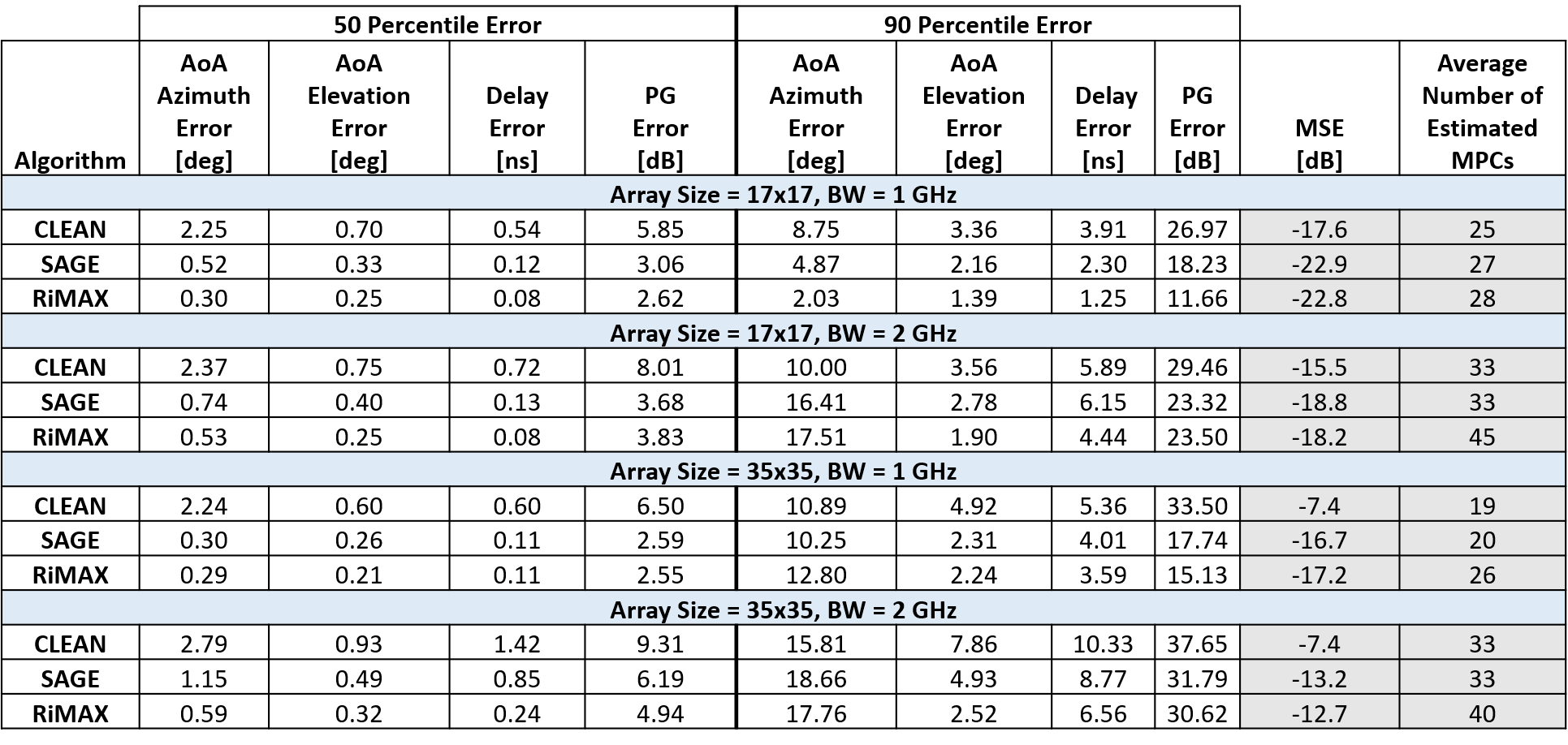}} 
\caption{\footnotesize{\sl The $50\,\%$ and $90\,\%$ errors in estimation of MPC parameters for three algorithms and for the different sounder configurations.}}
\label{fig:error_table}
\vspace{-4mm}
\end{table}

\vspace{-3mm}
\subsection{Real Measurement Results}
\label{sec:measurements}
\begin{figure}[hbt]
\vspace{-1mm}
\centering
\begin{tabular}{c}
\hspace{-2mm}
\includegraphics[width=3.4in]{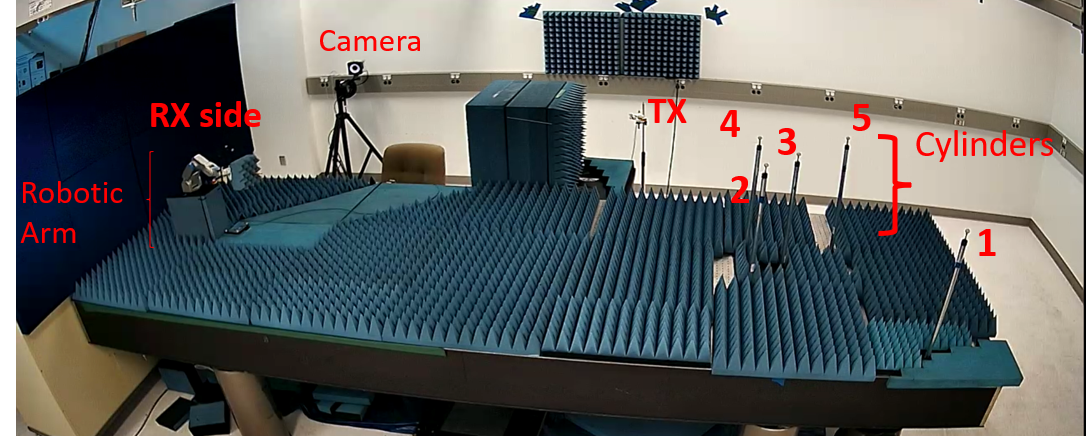} \\[-2mm]
 \footnotesize{(a)} \\
\includegraphics[width=2.0in]{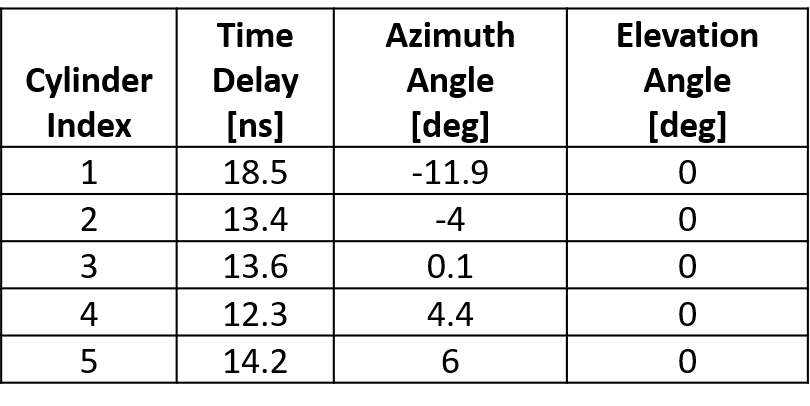} \\[-2mm]
 \footnotesize{(b)} 
\end{tabular}
\caption{\footnotesize{\sl (a) A photograph of the set up for the cylinder measurements. (b) The ground truth values of AZ, EL and delay for the 5 cylinders.}}
\label{fig:cyl_setup}
\vspace{-1mm}
\end{figure}
In this section, results are presented for the three algorithms to estimate the location of five cylinders based on real measurements collected by the SAMURAI sounder. The set up is shown in Fig.~\ref{fig:cyl_setup} which also includes the GT values of the AZ, EL and delay  for the cylinders. Five metal cylinders with diameter 2.5 cm are placed on an optical table. An optical tracking system reports the 3D locations of the cylinders and the TX and RX antennas with an accuracy of 0.01 mm, from which the GT values are computed. All cylinders have an EL angle of zero degree. Note that delays of cylinders 2 and 3 are within the resolution  at 2 GHz, and the AZ angles of cylinder pairs (2,3), (3,4) and (4,5) are within the angular resolution for the larger array size (35 $\times 35)$.
\begin{figure}[htb]
\begin{tabular}{cc}
\includegraphics[width=1.7in]{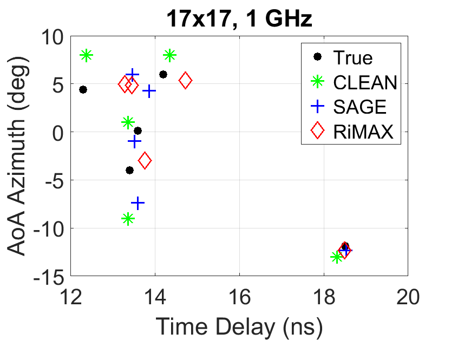}  & 
\hspace{-7mm} \includegraphics[width=1.7in]{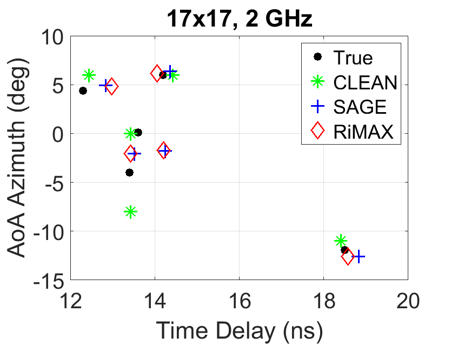} \\
\includegraphics[width=1.7in]{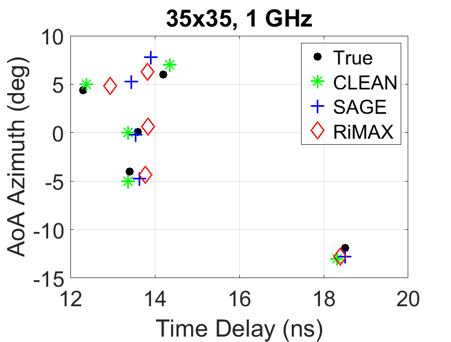}  & 
\hspace{-7mm} \includegraphics[width=1.7in]{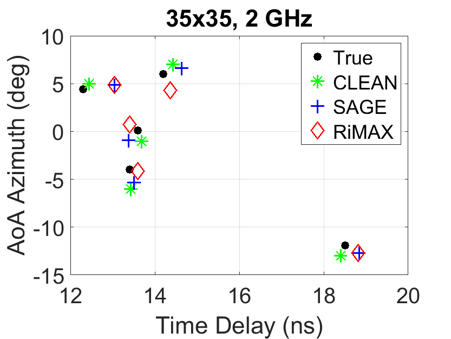} 
\end{tabular}
\caption{\footnotesize{\sl True AZ-delay values for the 5 cylinders and their estimates for the three algorithms for the four sounder configurations.}} 
\label{fig:cyl_gt_est}
\vspace{-1mm}
\end{figure} 
Fig.~\ref{fig:cyl_gt_est} shows the estimated values of AZ and delay, along with the GT values, for the four sounder configurations. Table~\ref{fig:meas_table} shows the estimation errors for all three algorithms and all four sounder configurations, for the 5 MPCs. The mean errors in AZ, EL and delay, averaged over the five cylinders are also included. All algorithms yield good estimates and the angular errors are within the sampling resolution for the array size. Overall, SAGE improves on CLEAN, and RiMAX yields the most accurate estimates in AZ. CLEAN yields the smallest delay errors but the errors for all algorithms are within or close to the sampling resolution for the sounder bandwidth. The ``0'' EL estimates for CLEAN are the closest grid point in the grid-based search.
\begin{table}[hbt]
\vspace{-1mm}
\centerline{\includegraphics[width=3.5in]{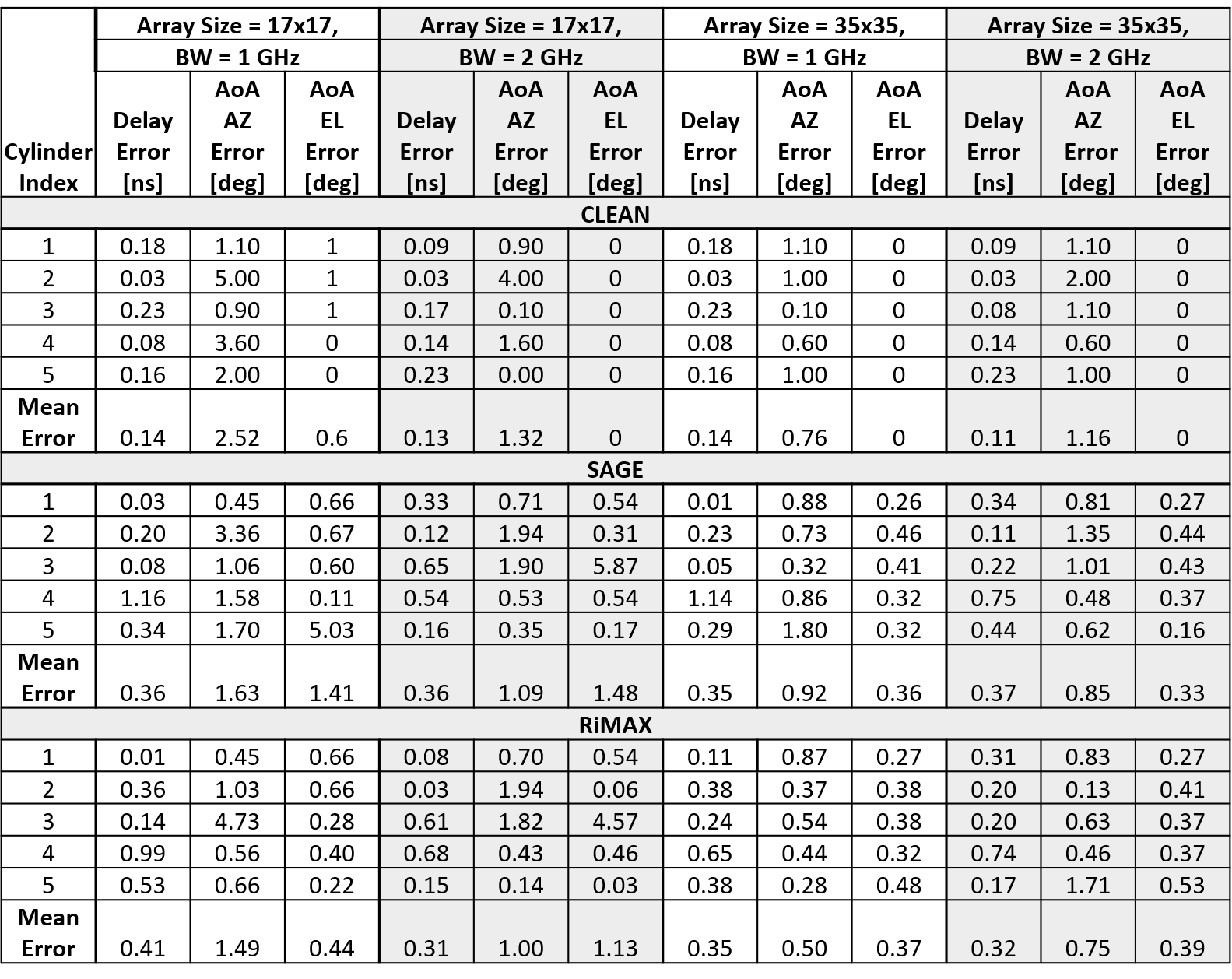}} 
\caption{\footnotesize{\sl A table showing the errors between estimated and true MPC parameters values for the three algorithms for the real measurements.}}
\label{fig:meas_table}
\vspace{-5mm}
\end{table}

\vspace{-3mm}
\section{Conclusions}
\label{sec:conc}
The new comprehensive mathematical framework developed in this paper builds on the initial results in \cite{sayeed_mpc:gcom20} (for CLEAN in an ``idealized'' noise-free case), to incorporate the effects of noise, non-ideal and cross-polar antenna element beampatterns, and multiple FoVs in both the system (channel and sounder) model as well as algorithm design. The new system model for this general realistic setting is leveraged for developing a new integrated framework encompassing three ML-based algorithms, with progressively higher complexity and performance, in the practically relevant single-snapshot setting: CLEAN, SAGE and RiMAX. CLEAN and SAGE are based on the purely specular MPC model, whereas RiMAX also accounts for the diffuse MPCs. The CLEAN algorithm is foundational as both SAGE and RiMAX get seeded by its output and utilize it in an iterative fashion.

The developed algorithms are evaluated using synthetic as well as real measurements. 
The synthetic sounder measurements generated from the NIST GT data represent one of the most complex propagation scenarios addressed in existing literature. The real measurements represent data collected from the SAMURAI sounder in a highly controlled environment with accurate measurements of the ground truth. The performance of algorithms is evaluated using two metrics: i) errors based on MPC association between estimated and GT parameters, and ii) reconstruction NMSE when no GT data is available. The evaluation is performed for two different array sizes and bandwidths to test the robustness of the algorithms.

Some comments on ``lessons learned'' are in order. First, while RiMAX generally gives the best performance, for both synthetic and measured data, the difference in performance of the three algorithms is not large.  Second, the results on the largest UPA size (35 $\times$ 35) and bandwidth (2 GHz) indicate that they represent the boundary of the widely used ``narrowband'' assumption \cite{brady:icc15}.
Third, the complexity and performance of the algorithms can vary depending on how the various algorithmic elements are implemented; see Sec.~\ref{sec:concepts}. For example, the LS update and path gain thresholding can serve as a powerful tool for pruning spurious MPC estimates. Similarly, checking consistency between the different hypothesized AZ values, driven by different UPA orientations, also aids in rejecting spurious MPCs. Finally, in implementations involving multiple algorithms, e.g., CLEAN + SAGE, or CLEAN + SAGE + RiMAX, the nature in which the procedure iterates between the different models is also important. For example, iterations across the SMC and DMC components could be done for each detected MPC, or after a certain number of MPCs have been detected. 

Computational complexity, driven by the high-dimensional nature of the problem, is also a significant consideration. In the results, the dimension of the spatio-temporal signal space varies from $N_o=28900$ ($57800$ for 2 GHz) for the 17 $\times$ 17 UPA to $N_o=122500$ ($245000$ for 2 GHz) for the 35 $\times$ 35 UPA. In particular, in RiMAX, the DMC update requires computing the inverse of the $N_o \times N_o$ matrix,  $\bR_{dan}$, over the entire search range of the 3D DMC parameters.
This is why the reported implementations of RiMAX, including this work, are limited to DMC modeling in the 1D delay domain, while ignoring spatio-temporal correlations. The estimation of a spatially and temporally variant DMC model is expected to improve the performance of RiMAX. Some initial work on spatial estimation is reported in \cite{kaske:10,kaske:15}. Implementing the algorithm in the beamspace promises very significant computational advantages by limiting the operations to a lower $D$-dimensional subspace, induced by a ``dominant'' thresholded version of the 3D PDP in AZ-EL-delay. All computations now occur in this lower operational dimension $D$, including matrix inversions. For example, for a 60 dB thresholding of the 3D PDP, only 4.5\,\% of the dimensions are retained ($N_o=122500$ reduces to $D=5493$) while retaining 99.7\,\% of the power.

Initial studies for relaxing the narrowband assumption~\cite{semper:wideband:23} suggest that the performance gains of RiMAX can be extended to scenarios with larger bandwidth and apertures, at the cost of higher complexity. Generally this requires the use of frequency-dependent array patterns. The results on addressing the ``beamsquint'' effect in \cite{brady:icc15} suggest an alternative approach for optimizing the performance versus complexity tradeoff by using multiple frequency-independent array patterns, as in the ``narrowband'' case, corresponding to multiple fixed beams centered around the true direction.

The framework and results reported in this paper open up several directions for future research, including: i) application of machine learning techniques for tuning the algorithmic parameters and for decision making in the execution of the algorithms; ii) extension of RiMAX to the full 3D AZ-EL-delay space, and iii) extension to sounders with large arrays and bandwidths. 

\vspace{-3mm}
\section{Acknowledgments}
\label{sec:ack}
The authors acknowledge the contributions of Jack Chuang (NIST) and Andreas Molisch (USC) in the initial discussions on the scope and approach of the project. 


\vspace{-3mm}

\bibliographystyle{IEEEtran}
\bibliography{bib/mpc_extraction}

\end{document}

%% file: akbar_defs_new.tex
\newcommand{\vc}{\mathrm{vec}}

\newcommand{\balpha}{\boldsymbol{\alpha}}

\newcommand{\snr}{\small \mbox{SNR}}

\newcommand{\mse}{\mathrm{MSE}}

\newcommand{\bmu}{\boldsymbol{\mu}}

\newcommand{\bA}{\boldsymbol{A}}

\newcommand{\bG}{\boldsymbol{G}}
\newcommand{\bH}{\boldsymbol{H}}
\newcommand{\bI}{\boldsymbol{I}}

\newcommand{\bR}{\boldsymbol{R}}

\newcommand{\bU}{\boldsymbol{U}}

\newcommand{\bW}{\boldsymbol{W}}

\newcommand{\ba}{\boldsymbol{a}}

\newcommand{\bg}{\boldsymbol{g}}
\newcommand{\bh}{\boldsymbol{h}}

\newcommand{\bw}{\boldsymbol{w}}
\newcommand{\bx}{\boldsymbol{x}}
\newcommand{\by}{\boldsymbol{y}}

\newcommand{\bzero}{\boldsymbol{0}}

\newcommand{\bVb}{{\tilde \bVb}}

\newcommand{\cO}{\mathcal{O}}

\newcommand{\cS}{{\mathcal S}}

\newcommand{\cCN}{\mathcal{CN}}



%% file: mpc_est_journal_twocol_final.bbl
\begin{thebibliography}{10}
\providecommand{\url}[1]{#1}
\csname url@samestyle\endcsname
\providecommand{\newblock}{\relax}
\providecommand{\bibinfo}[2]{#2}
\providecommand{\BIBentrySTDinterwordspacing}{\spaceskip=0pt\relax}
\providecommand{\BIBentryALTinterwordstretchfactor}{4}
\providecommand{\BIBentryALTinterwordspacing}{\spaceskip=\fontdimen2\font plus
\BIBentryALTinterwordstretchfactor\fontdimen3\font minus
  \fontdimen4\font\relax}
\providecommand{\BIBforeignlanguage}[2]{{%
\expandafter\ifx\csname l@#1\endcsname\relax
\typeout{** WARNING: IEEEtran.bst: No hyphenation pattern has been}%
\typeout{** loaded for the language `#1'. Using the pattern for}%
\typeout{** the default language instead.}%
\else
\language=\csname l@#1\endcsname
\fi
#2}}
\providecommand{\BIBdecl}{\relax}
\BIBdecl

\bibitem{ch_ams:brady_taps:12}
J.~Brady, N.~Behdad, and A.~Sayeed, ``Beamspace {MIMO} for millimeter-wave
  communications: System architecture, modeling, analysis and measurements,''
  \emph{IEEE Trans. on Antennas and Propagation}, pp. 3814--3827, July 2013.

\bibitem{ch_ams:sayeed_gcom:16}
A.~Sayeed and J.~Brady, ``Beamspace {MIMO} channel modeling and measurement:
  Methodology and results at {28 GHz},'' \emph{IEEE Globecom Workshop on
  Millimeter-Wave Channel Modeling}, Dec. 2016.

\bibitem{sayeed_mpc:gcom20}
A.~Sayeed, P.~Vouras, C.~Gentile, A.~Weiss, J.~Quimby, Z.~Cheng, B.~Modad,
  Y.~Zhang, C.~Anjinappa, F.~Erden, O.~Ozemir, R.~Muller, D.~Dupleich, H.~Niu,
  D.~Michelson, and A.~Hughes, ``A framework for developing algorithms for
  estimating propagation parameters from measurements,'' in \emph{IEEE Global
  Communications Conf.}, Taipei, Taiwan, Dec. 2020, pp. 1--6.

\bibitem{nist_twc:20}
C.~Gentile, A.~Molisch, J.~Chuang, D.~Michelson, A.~Bodi, A.~Bhardwaj,
  O.~Ozdemir, W.~Khwaja, I.~Guvenc, Z.~Cheng, F.~Rottenberg, T.~Choi,
  R.~M\"{u}ller, N.~Han, and D.~Dupleich, ``Methodology for benchmarking
  radio-frequency channel sounders through a system model,'' \emph{IEEE Trans.
  Wireless Communications}, pp. 6504--6519, 2020.

\bibitem{brady:icc15}
J.~Brady and A.~Sayeed, ``Wideband communication with high dimensional arrays:
  New results and transceiver architectures,'' in \emph{IEEE Int’l. Conf.
  Commun. Workshop (ICCW)}, June 2015.

\bibitem{nist_mpc:19}
C.~Lai, R.~Sun, C.~Gentile, P.~Papazian, J.~Wang, and J.~Senic, ``Methodology
  for multipath-component tracking millimeter-wave channel modeling,''
  \emph{IEEE Trans. Antenna and Propagation}, Mar. 2019.

\bibitem{riu}
R.~Feng, Y.~Liu, J.~Huang, J.~Sun, C.-X. Wang, and G.~Goussetis, ``Wireless
  channel parameter estimation algorithms: Recent advances and future
  challenges,'' \emph{China Communications}, vol.~15, no.~5, pp. 211--228,
  2018.

\bibitem{krim}
H.~Krim and M.~Viberg, ``Two decades of array signal processing research: the
  parametric approach,'' \emph{IEEE Signal Processing Magazine}, vol.~13,
  no.~4, pp. 67--94, 1996.

\bibitem{gaillot}
D.~Gaillot, E.~Tanghe, P.~Stefanut, W.~Joseph, M.~Lienard, P.~Degauque, and
  L.~Martens, ``Accuracy of specular path estimates with {ESPRIT} and {RiMAX}
  in the presence of measurement-based diffuse multipath components,''
  \emph{Proc. of the 5th European Conf. on Antennas and Propagation (EUCAP)},
  pp. 3619--3622, 2011.

\bibitem{tanghe}
E.~Tanghe, L.~Martens, W.~Joseph, D.~P. Gaillot, M.~Liénard, and P.~Degauque,
  ``Specular path estimation errors with {ESPRIT}, {SAGE}, and {RiMAX} in the
  presence of dense multipath,'' \emph{2012 IEEE-APS Topical Conf. on Antennas
  and Propagation in Wireless Communications}, pp. 315--316, 2012.

\bibitem{feng}
R.~Feng, Y.~Liu, J.~Huang, J.~Sun, and C.-X. Wang, ``Comparison of {MUSIC},
  unitary {ESPRIT}, and {SAGE} algorithms for estimating 3d angles in wireless
  channels,'' \emph{2017 IEEE/CIC Intl. Conf. on Commun. in China}, pp. 1--6,
  2017.

\bibitem{yin}
X.~Yin, L.~Ouyang, and H.~Wang, ``Performance comparison of {SAGE} and {MUSIC}
  for channel estimation in direction-scan measurements,'' \emph{IEEE Access},
  vol.~4, pp. 1163--1174, 2016.

\bibitem{damico}
A.~A. D'Amico, M.~Morelli, and L.~Sanguinetti, ``{DOA} estimation in the uplink
  of multicarrier {CDMA} systems,'' \emph{EURASIP Journal on Wireless
  Communications and Networking}, vol. 2008, pp. 1--9, 2007.

\bibitem{yao}
W.~Li, W.~Yao, and P.~J. Duffett-Smith, ``Comparative study of joint {TOA/DOA}
  estimation techniques for mobile positioning applications,'' in \emph{2009
  6th IEEE Consumer Communications and Networking Conf.}, 2009, pp. 1--5.

\bibitem{naitong}
W.~Yang and Z.~Naitong, ``A new multi-template {CLEAN} algorithm for {UWB}
  channel impulse response characterization,'' in \emph{2006 Intl. Conf. on
  Communication Technology}, 2006, pp. 1--4.

\bibitem{tschudin}
M.~Tschudin, C.~Brunner, T.~Kurpjuhn, M.~Haardt, and J.~Nossek, ``Comparison
  between unitary {ESPRIT} and {SAGE} for 3-d channel sounding,'' \emph{1999
  IEEE 49th Vehicular Technology Conf.}, vol.~2, pp. 1324--1329 vol.2, 1999.

\bibitem{ch_ams:bonek:01}
P.~M. Steinbauer, A.~F. Molisch, and E.~Bonek, ``The double-directional radio
  channel,'' \emph{IEEE Antennas and propagation Magazine}, vol.~43, no.~4, pp.
  51--63, 2001.

\bibitem{ch_ams:cost2100:12}
L.~Liu, C.~Oestges, J.~Poutanen, K.~Haneda, P.~Vainikainen, F.~Quitin,
  F.~Tufvesson, and P.~Doncker, ``The {COST} 2100 {MIMO} channel model,''
  \emph{IEEE Trans. Wireless Commun.}, vol.~19, no.~6, pp. 92--99, 2012.

\bibitem{ch_ams:sayeed:02}
A.~M. Sayeed, ``Deconstructing multi-antenna fading channels,'' \emph{IEEE
  Trans. Signal Processing}, vol.~50, no.~10, pp. 2563--2579, Oct. 2002.

\bibitem{ch_ams:sayeed_book:08}
A.~Sayeed and T.~Sivanadyan, \emph{Wireless Communication and Sensing in
  Multipath Environments using Multi-antenna Transceivers}.\hskip 1em plus
  0.5em minus 0.4em\relax Handbook on Array Processing and Sensor Networks (K.
  J. R. Liu and S. Haykin, Eds.), IEEE-Wiley, 2010.

\bibitem{eadf:04}
M.~Landmann and G.~Del~Galdo, ``Efficient antenna description for {MIMO}
  channel modelling and estimation,'' in \emph{7th European Conf. on Wireless
  Technology, 2004.}, 2004, pp. 217--220.

\bibitem{clean:74}
J.~A. H\"{o}gbom, ``Aperture synthesis with a non-regular distribution of
  interferometer baselines,'' \emph{Astron. Astrophys. Supp}, vol.~15, pp.
  417--426, 1974.

\bibitem{sage:94}
J.~A. {Fessler} and A.~O. {Hero}, ``Space-alternating generalized
  expectation-maximization algorithm,'' \emph{IEEE Trans. on Signal
  Processing}, vol.~42, no.~10, pp. 2664--2677, 1994.

\bibitem{ch_ams:richter:05}
A.~Richter, ``{E}stimation of {R}adio {C}hannel {P}arameters,'' \emph{PhD
  Thesis, TU Ilmenau}, 2005.

\bibitem{mp_mallat:93}
S.~G. {Mallat} and {Z. Zhang}, ``Matching pursuits with time-frequency
  dictionaries,'' \emph{IEEE Trans. on Signal Processing}, vol.~41, no.~12, pp.
  3397--3415, 1993.

\bibitem{Brewer}
J.~W. Brewer, ``Kronecker products and matrix calculus in system theory,''
  \emph{IEEE Trans. Circ. and Syst.}, vol.~25, no.~9, pp. 772--781, Sep. 1978.

\bibitem{STEWART:97}
M.~Stewart, ``Cholesky factorization of semidefinite {T}oeplitz matrices,''
  \emph{Linear Algebra and its Applications}, vol. 254, no.~1, pp. 497--525,
  1997, proceeding of the Fifth Conference of the International Linear Algebra
  Society.

\bibitem{kaske:10}
M.~Käske and R.~Thomä, ``Validation of estimated dense multipath components
  with respect to antenna array calibration accuracy,'' in \emph{Proc. of the
  4th European Conf. on Antennas and Propagation}, 2010, pp. 1--3.

\bibitem{kaske:15}
------, ``Maximum-likelihood based estimation of angular parameters of
  dense-multipath-components,'' in \emph{Proc. of the 9th European Conf. on
  Antennas and Propagation}, 04 2015, pp. 1--3.

\bibitem{semper:wideband:23}
S.~Semper, M.~Döbereiner, C.~Steinmetz, M.~Landmann, and R.~S. Thomä, ``High
  resolution parameter estimation for wideband radio channel sounding,''
  \emph{IEEE Trans. on Antennas and Propagation}, pp. 1--1, 2023.

\end{thebibliography}
